\documentclass{jpp}

\pdfoutput=1

\usepackage{physics}
\usepackage{color}
\usepackage{ifthen}
\usepackage[toc,page]{appendix}
\usepackage{dsfont}
\usepackage{url}
\usepackage[round]{natbib}

\usepackage{graphicx}
\usepackage{epstopdf, epsfig}
\usepackage{hyperref}
\usepackage{amsmath,amsfonts,amssymb}
\usepackage{subfigure}
\usepackage{bm}

\newcommand{\code}{\texttt}

\def\pindex{p}
\def\qindex{q}
\def\Tparam{\hbar}
\def\Lparam{L}
\def\Kparam{K}
\def\kparam{k}

\def\Hqsm{{\mathsfbi H}_{\rm QSM}}    
\def\Uqsm{{\mathsfbi U}_{\rm QSM}}    
\def\Upot{{\mathsfbi U}_{\rm pot}}    
\def\Ukin{{\mathsfbi U}_{\rm kin}}    
\def\Uqft{{\mathsfbi U}_{\rm QFT}}    
\def\Uphase{{\mathsfbi U}_{\rm phase}}    
\def\Up{{\mathsfbi U}_{\rm p}}
\def\matrho{{\bm \rho}}
\def\matsigma{{\bm \sigma}}
\def\matH{{\mathsfbi H}}
\def\matL{{\mathsfbi L}}
\def\matU{{\mathsfbi U}}

\newcommand{\newtext}[1]{{\textcolor{black}{#1}}}  
\newcommand{\changedtext}[1]{{\textcolor{black}{#1}}}  
\newcommand{\addedtext}[1]{{\textcolor{black}{#1}}}  

\title{Impact of dynamics, entanglement, and Markovian noise on the fidelity of few-qubit digital quantum simulation}

\begin{document}

\include{macros}

\author{M. D. Porter\aff{1,2}\corresp{\email{mdport@sandia.gov}} and I. Joseph\aff{1}}

\affiliation{
	\aff{1}Fusion Energy Sciences Program, Lawrence Livermore National Laboratory, Livermore, CA, USA
	\aff{2}Quantum Algorithms and Applications Collaboratory (QuAAC), Sandia National Laboratories, Albuquerque, NM, USA
}

\maketitle

\begin{abstract}
	Quantum algorithms have been proposed to accelerate the simulation of the chaotic dynamical systems that are ubiquitous in the physics of plasmas. Quantum computers without error correction \newtext{might} even use noise to their advantage to calculate the Lyapunov exponent by measuring the  Loschmidt echo fidelity decay rate. For the first time, digital Hamiltonian simulations of the quantum sawtooth map, performed on the {IBM-Q} quantum hardware platform, show that the fidelity decay rate of a digital quantum simulation increases \newtext{during the transition from dynamical localization to chaotic diffusion} in the map.
	 The observed error per \code{CNOT} gate increases by $1.5\times$ as the dynamics varies from localized to diffusive, while only changing the phases of virtual \code{RZ}  gates and keeping the over-all gate count constant.  A gate-based Lindblad noise model that captures the effective change in relaxation and dephasing errors during gate operation qualitatively explains the effect of dynamics on fidelity as being due to the localization and entanglement of the states created. Specifically, highly delocalized states \newtext{that are entangled with random phases} show an increased sensitivity to dephasing and, on average, a similar sensitivity to relaxation as localized states. In contrast, delocalized unentangled states show an increased sensitivity to dephasing but a lower sensitivity to relaxation. This gate-based Lindblad model is shown to be a useful benchmarking tool by estimating the effective Lindblad coherence times during \code{CNOT} gates and finding a consistent $2\text{--}3\times$ shorter $T_2$ time than reported for idle qubits.
	 Thus, the interplay of the dynamics of a simulation with the noise processes that are active can strongly influence the overall fidelity decay rate.
\end{abstract}

\section{Introduction}

\def\Bk{{\bf k}}
\def\Bv{{\bf v}}

\subsection{Motivation}
Quantum computers may eventually become indispensable to scientific computing due to their ability to accelerate many calculations of interest.
 Because the physics of plasmas crucially relies on understanding how the motion of charged particles responds to electromagnetic fields {and how the fields, in turn, respond to the particles}, understanding how quantum computers can be used to accelerate the simulation of complex nonlinear dynamics is of great importance to the field. 
 Plasma transport strongly depends on whether the particle motion possesses adiabatic invariants, such as the magnetic moment, or whether the invariants are destroyed and the motion is chaotic. While the former is the basis for particle confinement, the latter is used for heating plasmas to fusion relevant temperatures. 
 {In many situations of interest}, the dynamics of the fields are often chaotic or even turbulent, which enhances the transport of particles, heat, and momentum through the plasma.  
 Thus, a key research direction for quantum algorithms for plasma physics is to accelerate the simulation of the nonlinear, chaotic, and turbulent dynamics of plasmas \citep{joseph2023quantum}.  

\newtext{
Interactions between particles and fields in a plasma are often classified into wave-particle and wave-wave interactions. Wave-particle interactions refer to the change in the trajectories of particles due to the electromagnetic fields, the change in the trajectories of wave packets due to scattering from particles, and the resulting exchanges of energy and momentum \citep{davidson2012methods, nicholson1983plasma}.  
The electromagnetic forces generate evolution of the particle distribution function (PDF) in phase space, and, there are analogous processes that cause wave packets to evolve in wavenumber, $\Bk$, and frequency, $\omega$, space.  
Wave-particle interactions such as Landau damping, quasilinear theory, plasma echos, and induced scattering are ubiquitous.
Wave-wave interactions refer to nonlinear interactions between  wave packets of different types, which can, in fact, be mediated through the particles. 
This includes calculations of wave-wave scattering process, as well as both strong and weak turbulence theory \citep{zakharov2012kolmogorov, nazarenko2011wave}.
At a fundamental level, all of these processes can be viewed as the evolution of a plasma as a nonlinear dynamical system. 
}

\def\Efield{E}
\newtext{
For small amplitude waves, the effects of nonlinear interactions can be understood through an expansion in wave amplitude.  The linear response for wave-particle interactions is largest at a resonance between the particle velocity and the phase velocity of the wave, $\omega=\Bk\cdot\Bv$, where the phase velocity, $ \omega/k$, matches the particle velocity $\Bv$.  If the PDF decays in velocity at the location of the resonance relative to the bulk, then the waves experience Landau damping. 
The opposite case is unstable and inverse Landau damping causes the wave amplitude to grow. 
For finite wave amplitude, $\Efield$, particles are trapped in an island in phase space of finite width, $\delta v=\sqrt{2 \Efield }$, for a particle of unit charge and mass.  
When multiple waves are present with different frequencies and wavenumbers, resonances can occur at the phase velocity of the nonlinear beat wave, $\sum_i \omega_i=\sum_i \Bk_i \cdot \Bv$, which generalizes the resonance condition for wave-wave interactions.
Once the islands generated by waves at different phase velocities begin to overlap, i.e. above the Chirikov criterion  \citep{chirikov1979universal}, there is a transition to chaotic particle motion, where particles effectively diffuse in momentum space.  Quasilinear theory derives the effective diffusion that particles experience due to the wave power spectrum and the transfer of energy-momentum between waves and particles.  
In fact, this transition to chaos is generic for any dynamical system.
However, in the chaotic regions of phase space, the true motion is rather complex and accurate numerical simulations are required to determine the evolution of the PDF.
}

\newtext{
 In general, plasma simulations must be addressed with computational approaches for solving the relevant partial differential equations (PDEs), e.g. the Vlasov-Poisson or Vlasov-Maxwell system.  
 This requires both an accurate calculation of the particle orbits in the force-field generated by the waves and an accurate calculation of the response of the waves to the charge and current density of the particles. 
Due to the high-dimensional phase space, solving for the motion of the particle PDF is computationally demanding and is the largest computational overhead in today's kinetic codes, both for Lagrangian (particle-in-cell) and Eulerian (e.g. finite volume or finite element) approaches.  
Moreover, self-consistent calculations are highly demanding precisely because plasma evolution often leads to chaotic and turbulent dynamics that require extremely high spatial resolution for accurate results.  
}

\newtext{
The power of quantum computers to apply useful operations to high dimensional Hilbert spaces, essentially in parallel, implies that it may one day be possible to accelerate calculations of wave-particle interactions. 
In fact, we recently proposed the Koopman-von Neumann approach \citep{joseph2020koopman, joseph2023quantum} for using quantum computers to evolve all trajectories at once in an efficient manner, potentially leading to exponential speedup for the evolution of the PDF.
In this work, we shift focus from designing algorithms for future fault-tolerant quantum computers to understanding how such algorithms perform in practice on one of today's hardware platforms.   
In order to make the problem more tractable, we consider simulating a toy model of wave-particle interactions by simulating the evolution of the PDF in  a fixed electrostatic potential. 
Another paper in this special issue \citep{shi2024simulating} tests the ability of present-day quantum computers to simulate a toy model of the chaotic dynamics of wave-wave interactions as a proxy for nonlinear partial differential equations of interest to plasma physics. 
}

\newtext{
However, the present era of quantum computing has been dubbed the Noisy Intermediate-Scale Quantum (NISQ) era
because present-day quantum computers are  limited in the fidelity of the basic gate operations as well as in the overall number of gates (gate depth) that can be applied coherently.  
The presence of errors in the calculation requires some type of error-characterization and mitigation strategy, and, without error-correction, it is not yet possible to perform the high-precision calculations necessary for plasma simulation.  
Our conclusion is that it is important to control the type and strength of various noise processes in order to obtain accurate results.
}

\subsection{\newtext{Quantum Maps}}
 {
 The first step in simulating wave-particle interactions 
 is to compute how the particle trajectories changes in response to the fields.
 }
In the early days of plasma research, physicists explored the chaotic dynamics of the particle motion through the study of discrete time dynamical systems called nonlinear maps \citep{lichtenberg1992regular}.
Relatively simple {nonlinear} maps, such as the Chirikov standard map \citep{chirikov1979universal}, generate chaotic motion by breaking time invariance with a series of periodic kicks in time. 
 Nonlinear maps are both simpler and more accurate to study than nonlinear differential equations because they avoid numerical integration in time. Hence, they have no truncation errors associated with the discrete approximation of the integrals and, perhaps more importantly, they require far fewer operations per time step, so that numerical errors associated with finite precision arithmetic  are kept to a minimum.   Thus, nonlinear maps can be computed very efficiently and allow one to explore the rich structure of the chaotic dynamics of natural plasma processes such as the heating of magnetized plasmas by cyclotron waves  \citep{lichtenberg1992regular}.  
 Although the dynamics is controlled by a small set of parameters, it usually displays rich structural properties due to the multitude of bifurcations in the number and types of fixed points of the mapping as the parameters are varied \citep{guckenheimer2013nonlinear}.

\begin{figure}
	\centering
	\includegraphics[trim={0 0 0 8mm}, clip, width=120mm]{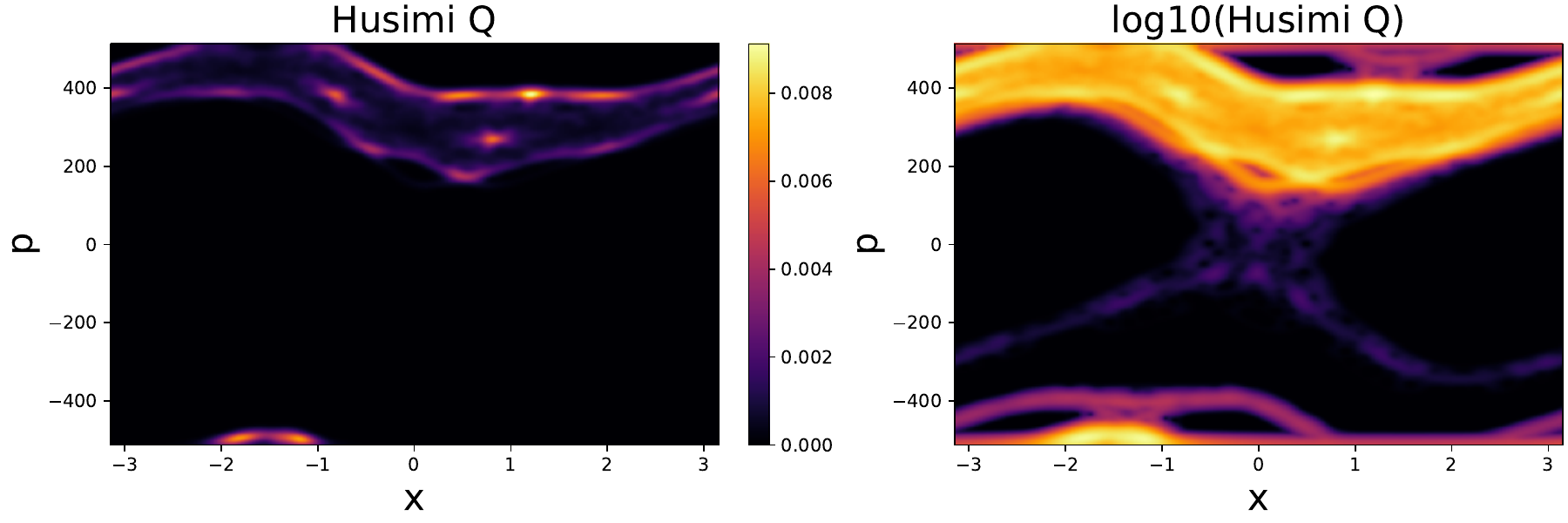} \\
	\includegraphics[trim={0 0 0 8mm}, clip, width=120mm]{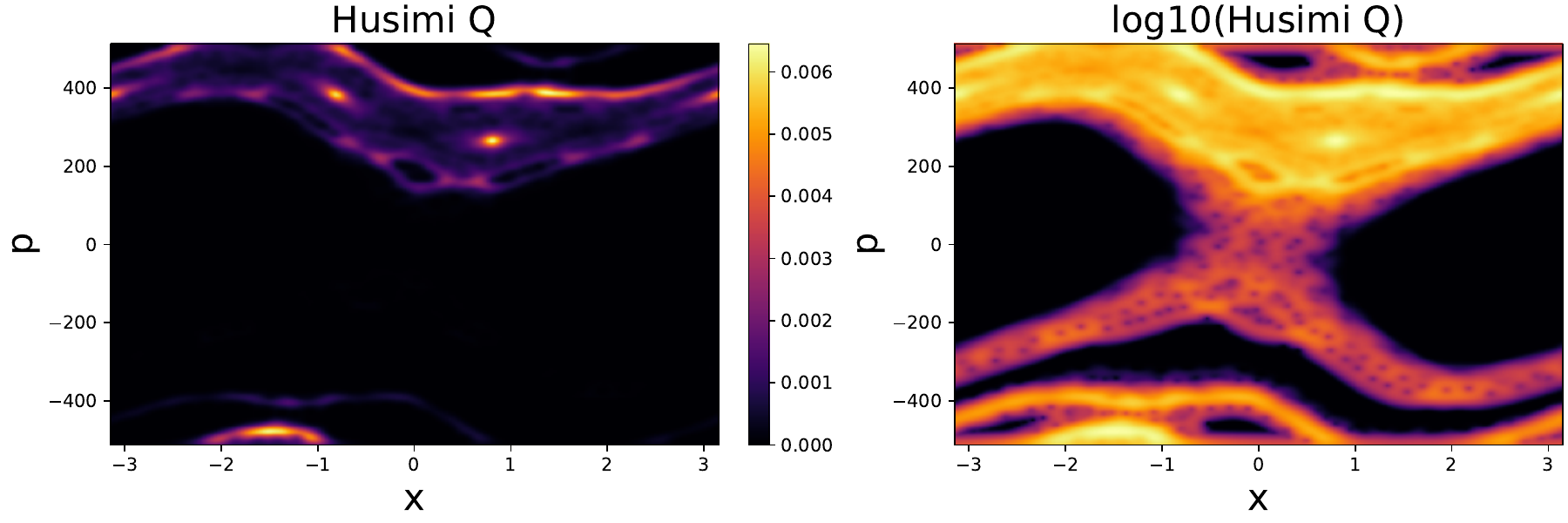} \\
	\includegraphics[trim={0 0 0 8mm}, clip, width=120mm]{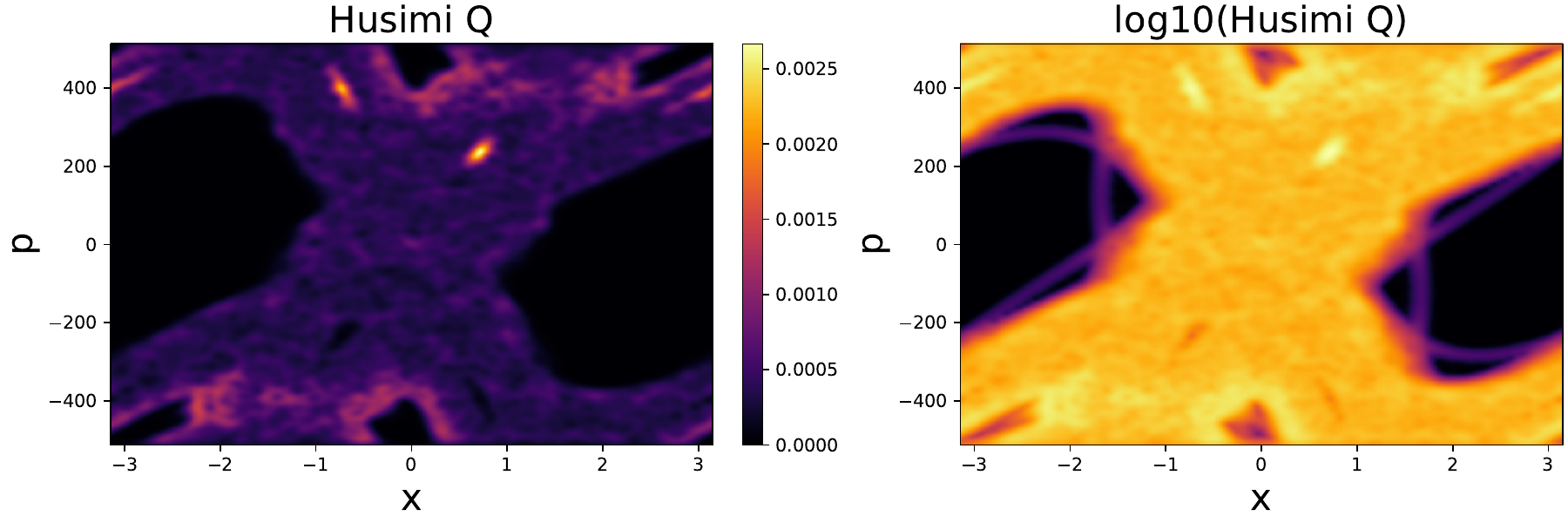}
	\caption{Husimi-Q quasiprobability distribution for the quantum standard map (same decomposition as Eq.~\ref{eq:QSM_def}, but with the potential in Eq.~\ref{eq:H_QSM} 
	modified to $K(1-\cos\hat{\theta})$ starting from initial condition $p=3N/8$ ($J=3\pi/4$) for 10 qubits.  
	The map is evolved for 1000 time steps and then the final probability distribution is averaged over the last 50 time steps.
	Parameters: $L=1$ and (top) $K=0.95$ below the destruction of last KAM surface; (middle) $K=1.0$ above the destruction of last KAM surface, (bottom) $K=1.5$ chaotic diffusive regime. \newtext{Calculations performed on a classical computer without noise.}}
	\label{fig:qstd_map}
\end{figure}

{
While there are formally exact methods for developing quantum algorithms that simulate classical dynamics \citep{joseph2020koopman, liu2021efficient, joseph2023quantum}, for small system sizes the method of quantization is cheaper in terms of resource requirements, i.e. number of qubits and quantum gates.  A symplectic nonlinear map can be \emph{quantized} by embedding the dynamics within a unitary transformation {in a manner} that reproduces the classical dynamics in the semiclassical limit \citep{benenti2004quantum}. While achieving the semiclassical limit is challenging classically, the exponential memory resources of a quantum computer and the ability to simulate superpositions {efficiently} make this approach feasible quantumly. }
{
For example, the quantized version of the Chirikov standard map, the prototypical example of chaotic particle motion in response to a nonlinear wave, is shown in Fig.~\ref{fig:qstd_map}.
The map is run starting from an initial condition localized in momentum space that then explores the accessible phase space over time.
\addedtext{One can clearly see a large scale island in the center of the figure, whose width scales as $ K^{1/2}$, as expected.
The fact that the map is kicked periodically in time also generates islands at other phase velocities and chaotic motion occurs near the regions where islands overlap.
}
As the map parameter $K$ varies across the point at which the last Kolmogorov-Arnold-Moser (KAM) surface is destroyed, (estimated as $K\simeq0.971635406$ using Greene's method  \citep{lichtenberg1992regular}), the wavefunction begins to diffuse over the entire phase space.
Thus, simulating the quantized evolution can be thought of as a \emph{quantum walk algorithm} that accelerates the exploration of and averaging over the accessible chaotic region \citep{joseph2023quantum, di2016discrete}.
}

{
Simulation of the exact quantum dynamics \citep{brodin2022quantum}, including quantum wave-particle interactions \citep{misra2022wave-particle}, is of great scientific interest, but challenging using classical computers due to the memory and time required to directly simulate the exponentially large Hilbert space for the fully quantized system.  
Hence, quantum simulation using quantum computers is one of the main applications of interest for achieving near term quantum advantage \citep{babbush2021focus}. Simulating chaotic dynamics is of particular interest due to the provable difficulty of simulating chaos. Recent claims of quantum supremacy relied on the difficulty of simulating chaotic quantum circuits \citep{boixo2018characterizing, arute2019quantum} and among paths to quantum advantage simulating chaotic dynamics may be the most qubit-efficient \citep{babbush2021youtube}. Thus, direct simulation of the quantized system opens the pathway both for simulating the intrinsically quantum dynamics as well as for accelerating the simulation of classical dynamics.  
}

{
This work explores the quantum sawtooth map (QSM) as a prototypical point-example of both the classical and quantum plasma physics simulations  that quantum computers may one day accelerate.  The QSM is one of the cheapest possible maps to simulate \citep{benenti2001efficient} because it only depends quadratically on the momentum and the position, which significantly reduces the number of arithmetic operations per time step, both classically and quantumly.  This has led to a proposal to use the QSM as a benchmark problem to measure the ability of quantum algorithms to accelerate the simulation of dynamical systems \citep{benenti2004quantum}.   
{The result of evolving the QSM from a localized initial momentum state is shown in Fig~\ref{fig:qsaw_map}.
Again, one can clearly see a transition to the regime of chaotic diffusion as the map parameter increases (compare to the standard map Fig~\ref{fig:qstd_map}) .
}
}

Both {chaotic/diffusive dynamics} and dynamical localization, which is an intrinsically quantum effect, can be observed in the quantum sawtooth map.
{(Here, we use the term diffusive rather than chaotic because the dynamics are not clearly chaotic until a sufficient number of qubits is used { \citep{porter2022observability}}.)} 
The transition from diffusion to localization occurs when the ratio of diffusion strength to {effective Planck's constant ($\hbar$)} is small and quantum interference dominates. 

\subsection{Leveraging noise}
The 
NISQ devices that are available today provide a unique platform for exploring the dynamics of many-body quantum systems. However, the ubiquitous presence of interactions with the environment generates noise which adds complexity to the interpretation of the results. Depending on context, noise can affect the simulation dynamics in different ways. It may completely wash out the dynamics of interest or it can potentially be used to measure key signatures of the dynamics \citep{porter2022observability}.

An important feature of quantum transport simulations is that they can be performed in the presence of noise. In fact, the algorithm for calculating the Lyapunov exponent measures the fidelity decay rate of a Loschmidt echo experiment which crucially relies on noise being present as part of the calculation.  On future error-corrected quantum computers, one can carefully introduce effective noise sources into the calculation by design.  
For today's noisy intermediate scale quantum (NISQ) computers, one \newtext{might utilize specific types of hardware noise that may already be present to one's advantage}, as long as the strength of the noise and the strength of the Lyapunov exponent can be chosen to reside within a certain window in parameter space \citep{porter2022observability}. \newtext{The ability to characterize and control the types of noise present could enable this by leveraging tailored noise \citep{van2023probabilistic, guimaraes2023noise}.} Quantum hardware platforms with enough qubits and with the right magnitude of noise to perform such calculations may be available in the near future. 

Measuring the decay of fidelity can provide an exponentially efficient measure of the classical Lyapunov exponent \citep{benenti2002quantum, benenti2004quantum} and chaotic decoherence \citep{poulin2004exponential}. 
Measuring dynamical localization of classically chaotic systems may yield a similar speedup \citep{georgeot2001exponential}. 
The fidelities of quantized versions of classically chaotic Hamiltonian systems can decay differently (often faster) than for integrable dynamics \citep{peres1984stability, lysne2020small, porter2022observability}. 
Quantum chaos can magnify the effect of Trotter errors \citep{sieberer2019digital} while quantum localization can {reduce the upper bound on their impact} \citep{heyl2019quantum}.

\subsection{Summary of results}

In this work we ask the question: how do dynamics, entanglement, and noise impact the fidelity of the results? {We show experimentally that varying the dynamics of the quantum sawtooth map alters the fidelity decay rate.}
\cite{henry2006localization} and \cite{pizzamiglio2021dynamical} initiated the use of the quantum sawtooth map to characterize experimental noise by using the degree of localization as a test of device fidelity. We extend previous work by studying the phase transition from localization to diffusion and the Loschmidt echo fidelity throughout this transition, opening the door to probing the interaction between quantum map dynamics and experimental noise.

The main results are:
\changedtext{(1) We report on the first experimental evidence demonstrating that the Loschmidt echo fidelity of a digital quantum simulation decreases as the dynamics transitions from integrable to chaotic even as the gate count remains constant. 
(2) While this behavior was anticipated by previous studies, we found that the parametric noise models that others had explored cannot explain the experimental results. 
(3) Instead, this behavior can be phenomenologically explained by the fact that chaotic evolution creates \newtext{randomly entangled states, defined as having significant amplitude on all basis states with random relative phases on each,} that are more sensitive to dephasing and similarly sensitive to relaxation as localized states. This can be attributed to an increase both in superposition, which increases the effect of pure dephasing and reduces the effect of relaxation, and in entanglement with random phases, which increases the effect of relaxation. 
(4) Three noise models that we explore agree on a common value of the effective $T_2$ time during gate operation that, as is to be expected, is far shorter than IBM-Q's reported value for $T_{2E}$ from a Hahn echo experiment.}

\changedtext{We introduce a} gate-based Lindblad model that both captures the effect of dynamics on fidelity in a minimal physically-motivated model and results in an effective $T_2$ time similar to that from a Qiskit Aer model fit. In contrast, the parametric noise model often used by other authors {in the quantum maps literature} does not capture the correct form of the fidelity decay, and \newtext{randomized benchmarking (RB)} with depolarizing noise does not have the minimum two parameters required to describe a fidelity that depends on more than gate count.

\begin{figure}
	\centering
	\includegraphics[trim={0 0 0 8mm}, clip, width=120mm]{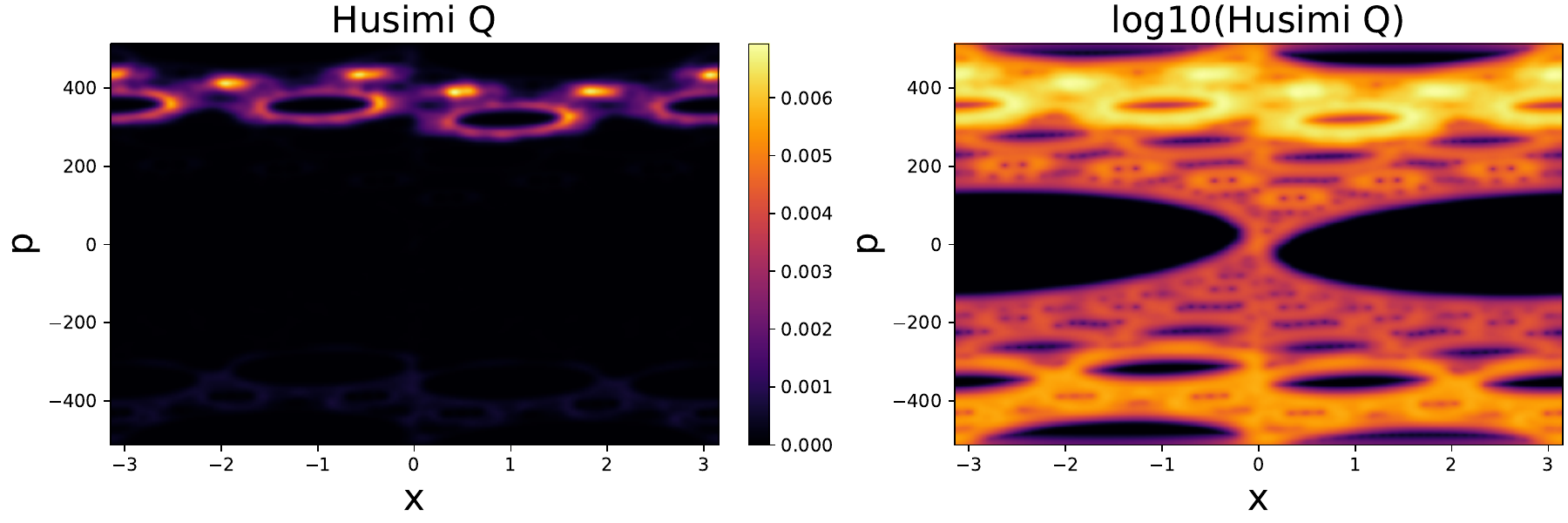}
	\includegraphics[trim={0 0 0 8mm}, clip, width=120mm]{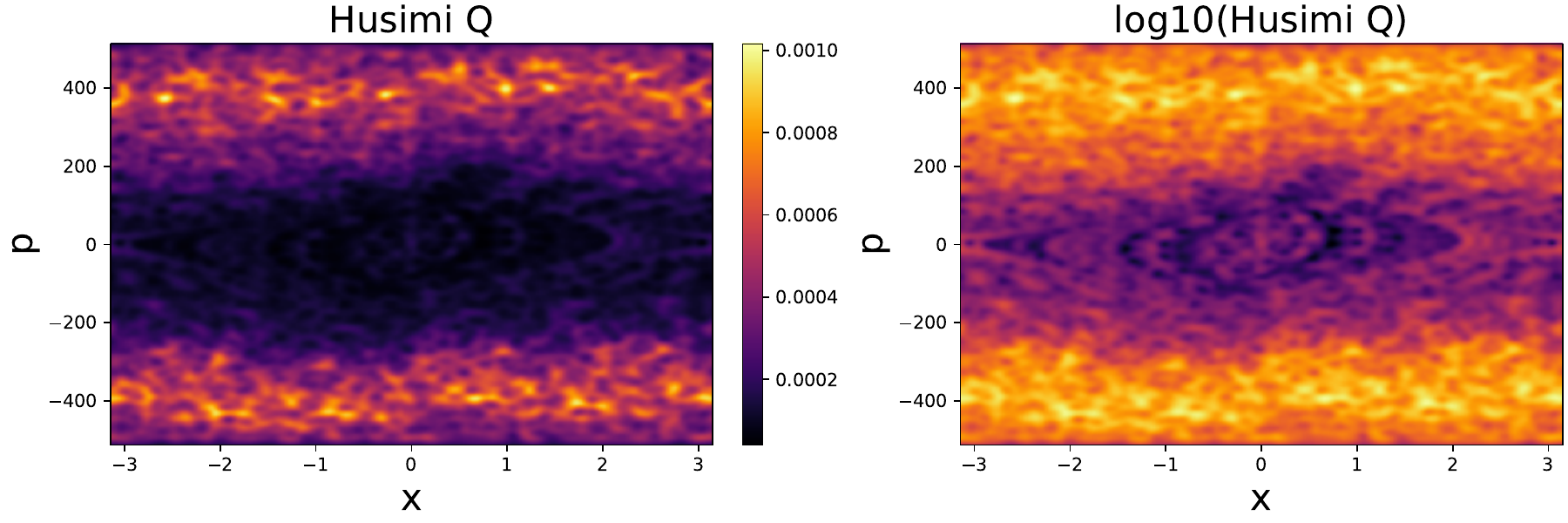}
	\includegraphics[trim={0 0 0 8mm}, clip, width=120mm]{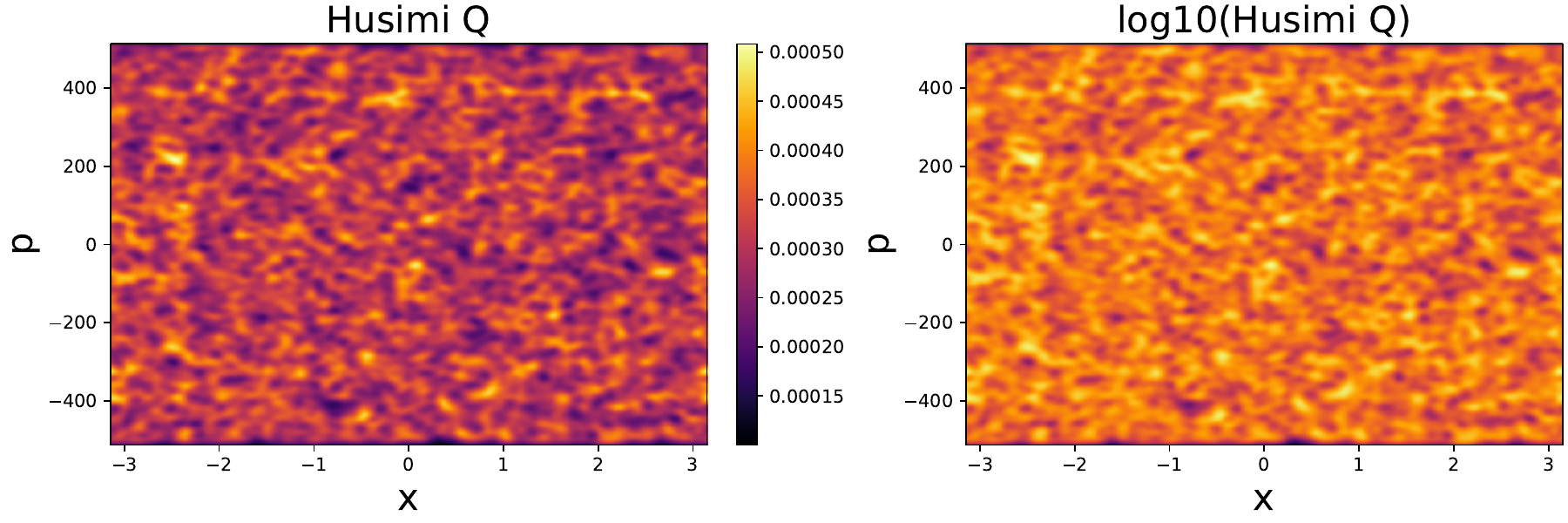}
	\caption{Husimi-Q quasiprobability distribution for the quantum sawtooth map (Eq.~\ref{eq:QSM_def}) for $n=10$ qubits, starting from initial condition $p=3N/8$ ($J=3\pi/4$). 
	The map is evolved for 1000 time steps and then the final probability distribution is averaged over the last 50 time steps.
	Parameters: $L=1$ and (top) $K=-0.1$ regular motion; (middle) $K=0.1$ anomalous diffusion; (bottom) $K=1.5$ chaotic diffusive regime. \newtext{Calculations performed on a classical computer without noise.}}
	\label{fig:qsaw_map}
\end{figure}

\subsection{Overview of contents }

{In Sec.~\ref{sec:qsm} the quantum sawtooth map is introduced, its conditions for dynamical localization are described, and its gate decomposition is given. In Sec.~\ref{sec:main_results} the experimental results are presented and compared to IBM-Q reported metrics.
In Sec.~\ref{sec:noise_models} noise models are described and fit to experimental data to extract effective decoherence times. {Appendix \ref{appendix:rationale} provides background rationale for the gate-based Lindblad model that is a primary tool in this section.} Section \ref{sec:lindblad} and Sec.~\ref{sec:lindblad_gate-based} consider \newtext{single-qubit} Lindblad noise models, Sec.~\ref{sec:ibm_sim_noise} considers the qiskit Aer noise model, and Sec.~\ref{sec:experiment_fit_lindblad} fits the three resulting models to the data.
Lastly Sec.~\ref{sec:conclusion} provides a summary of our results.} \changedtext{Discussion of the parametric noise model considered in previous work is relegated to Appendix~\ref{appendix:parameter_noise}.}

\section{Quantum sawtooth map}
\label{sec:qsm}

\subsection{Definition}

The quantum sawtooth map (QSM) is defined by the {dimensionless} time-periodic Hamiltonian
\begin{equation} 
	\label{eq:H_QSM}
	\Hqsm 
	=\frac{\hat{J}^2}{2} - \sum_n K \frac{\hat{\theta}^2}{2} \delta(t - n) \text{ for } \theta \bmod 2\pi
\end{equation}
which is derived {in Ref.~\citep{porter2022observability}} from the classical Hamiltonian by quantizing {in dimensionless $\hbar$} to get $\hat{J} = \Tparam \hat{\pindex}$ and discretizing to get $\hat{\theta} = 2 \pi \hat{\qindex}/N$ to yield momentum and position operators $\hat \pindex$ and $\hat \qindex$ respectively, with eigenvalues $-N/2 \le p, q < N/2$. The quantum evolution propagator over one period is then
\begin{flalign}
	\label{eq:QSM_def}
	\Uqsm &= \hat{\mathcal{T}} e^{-i \int_0^1 \Hqsm dt /\Tparam} = \Ukin \Upot \\
	\Upot(\hat \qindex) &= e^{i \kparam (\beta \hat \qindex)^2 /2}, \qquad 
	\Ukin(\hat \pindex) = e^{-i \Tparam {\hat \pindex}^2 /2} \nonumber
\end{flalign}
where $\hat{\mathcal{T}}$ is the time-ordering operator, $\kparam \equiv \Kparam/\Tparam$ is the quantum kicking parameter, and $\beta \equiv 2\pi/N$ for $N$ basis states. In the context of quantum computing $N=2^n$ for $n$ qubits. \newtext{Eq.~\ref{eq:QSM_def} is exact with no Trotter error due to the delta-function potential that is kicked periodically in time. In this instant the potential energy overwhelms the kinetic energy and so can be considered to occur at the beginning of each time step, entirely before the kinetic evolution. The whole} single-period propagator is often called a Floquet operator \newtext{with periodicity one \citep{rudner2020floquet}} since it uses a time-periodic Hamiltonian, but since it corresponds to a classical map it is also known as a quantum map. \newtext{(See the Conclusions \& Outlook section of \cite{mori2023floquet} for a review of Floquet theory in the context of open quantum systems.)}
Periodicity matching between the classical and quantum systems gives $\Tparam = 2\pi \Lparam /N$ for positive integer $\Lparam$ \citep{porter2022observability}.

Note a partial symmetry between the phases of $\Ukin$ and $\Upot$: $\Tparam=\Lparam* 2\pi/N$ while $-\kparam \beta^2 = -\Kparam/\Lparam* 2\pi/N$. Typically the qubits are mapped to the $\pindex$-basis, but mapping instead to the $\qindex$-basis would swap the roles of $\Lparam$ and $-\Kparam/\Lparam$. 

\subsection{Localization and initial conditions} \label{sec:localization}

{
The QSM is integrable when  $K=-4,-3,-2,-1, 0$, and, hence, the wavefunctions are localized for these cases.
For the intermediate region, $K\in(-4,0)$, the dynamics is not localized, but has zero Lyapunov exponent.
The dynamics of the QSM are chaotic for $K<-4$ and $K>0$.
For $0<K<1$, the wavefunction is chaotic, but it is in a regime of anomalously slow diffusion.
Figure~\ref{fig:qsaw_map} illustrates the difference in QSM dynamics between the zero Lyapunov case, $K=-0.1$, the anomalous slow diffusion case, $K=+0.1$, and the standard chaotic case, $K=1.5$. 
Since the focus of this paper is on chaotic dynamics, from now on only $K>0$ will be considered.
}

The presence of broken cantori in the classical system can slow diffusion at small $\Kparam$, so there are two regimes given by the classical diffusion coefficient
\begin{flalign}
	\label{eq:diff_coeff}
	D_\Kparam \approx \begin{cases}
		(\pi^2/3) K^2 \qquad \text{ for } \Kparam>1 \\
		3.3 \Kparam^{5/2} \quad \text{ for } 0<\Kparam<1 \\
	\end{cases}
\end{flalign}
which measures the rate of trajectories diffusing through the phase space \citep{benenti2001efficient}. When $D_{\Kparam}$ is small compared to $\Tparam^2$, the QSM is predicted to reach a steady state after the Heisenberg time that is exponentially localized around an initial momentum state $\ket{\pindex_0}$ as 
\begin{equation}
	P_\pindex = |\braket{\pindex}{\psi}|^2 \approx \frac{1}{\ell} e^{- \frac{2|\pindex-\pindex_0|}{\ell}}
\end{equation}
with localization length \citep{benenti2004quantum}
\begin{equation}
	\label{eq:loc_length}
	\ell \approx D_{\Kparam}/\Tparam^2.
\end{equation}
The Heisenberg time or ``break" time \citep{benenti2004quantum} is the time to resolve the energy levels, defined as the inverse mean energy level spacing \citep{shepelyansky2020ehrenfest,vsuntajs2020quantum}.

\begin{figure}
	\centering
	\includegraphics[width=80mm]{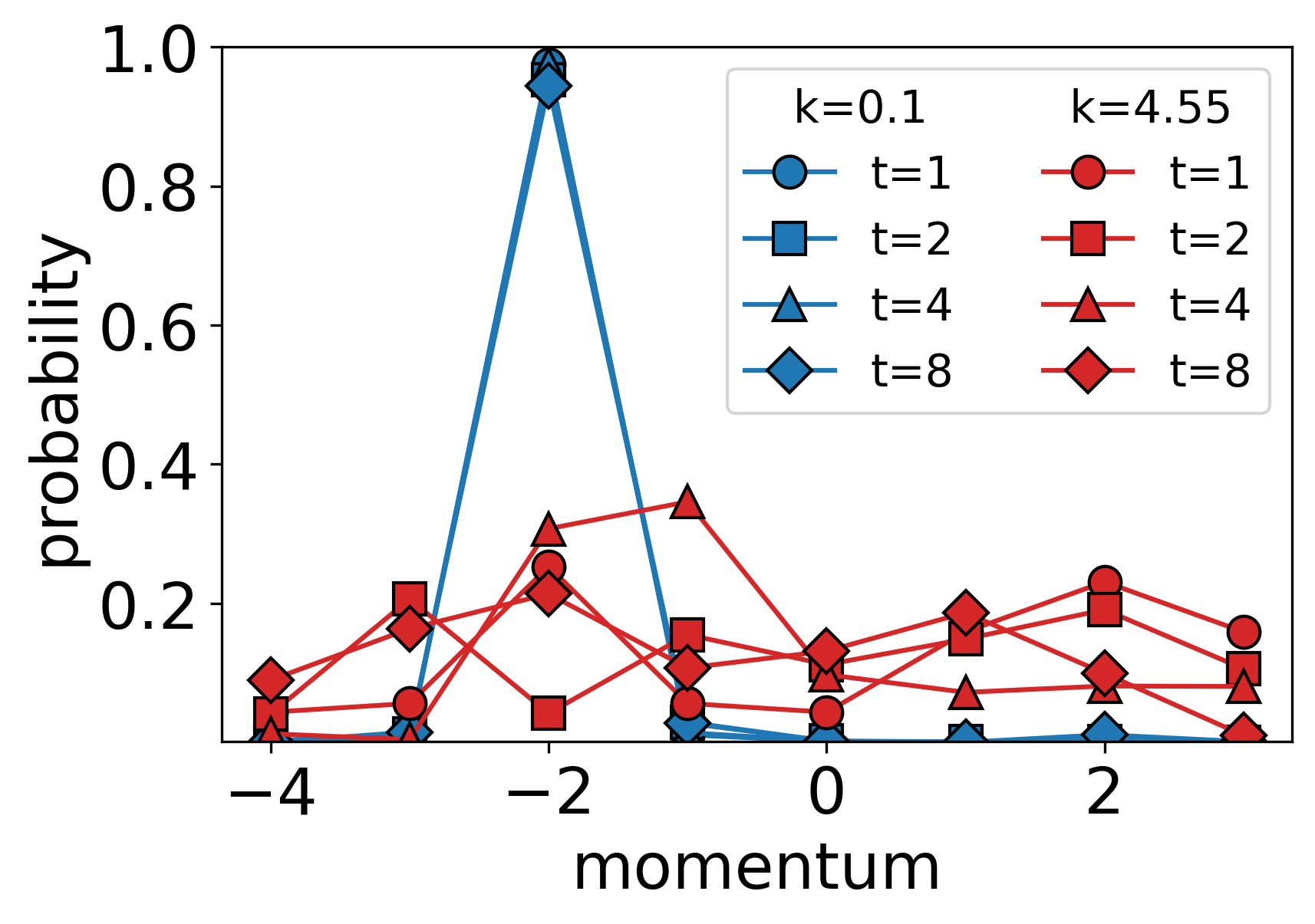}
	\caption{Exact noiseless simulations of the quantum sawtooth map (QSM), showing the localized case $k=0.1$ (blue) and the diffusive case $k=4.55$ (red) for $t=1,2,4,8$. {Initial state prepared in $\ket{\pindex=-2}$.} Parameters: $n=3\, (N=8), L=1; k=4K/\pi, k_\text{loc} \approx 1.87$. The horizontal axis corresponds to the vertical axis of Fig.\ref{fig:qsaw_map}, but at different $N$.}
	\label{fig:state_loc_dif} 
\end{figure}

Due to the periodicity and finite size of the system, localization only occurs if the localization length is small enough to have a global maximum at its central peak. This occurs when \citep{porter2022observability}
\begin{flalign}
	\label{eq:loc_condition}
	\kparam < \kparam_\text{loc}  &\equiv \begin{cases}
		\sqrt{\frac{3}{\ln(2) \pi^2} N} \approx 0.66 N^{1/2} \qquad\qquad\qquad\quad \text{ for } \Kparam>1 \\
		\left(\frac{1}{3.3 \sqrt{2\pi} \ln(2)} \frac{N^{3/2}}{\Lparam^{1/2}} \right)^{2/5} \approx 0.50 N^{3/5} \Lparam^{-1/5} \text{ for } 0<\Kparam<1
	\end{cases} \\ 
	&= \max(0.66 N^{1/2} , 0.50 N^{3/5} \Lparam^{-1/5})  \nonumber
\end{flalign}
with diffusion occuring otherwise. These two dynamical regimes are demonstrated in Fig.~\ref{fig:state_loc_dif}.

When localization is strong ($\ell \ll N$) the above formula for average $\ell$ does not uniformly apply to all initial conditions, with $\ell$ instead depending on initial condition in a manner dependent on $\Lparam$. Less localized clusters of momentum eigenstates appear at $\Lparam$ equally spaced locations in momentum space, reducing $\ell$ for initial momentum states in the vicinity of these groups. Once $\Lparam \sim N$ this transitions to more uniform $\ell$ with just $\ket{\pindex=0}$ strongly localized. This is how Refs.~\citep{henry2006localization, pizzamiglio2021dynamical} used $N=8, \Lparam=7$ to obtain strong localization at $\ket{\pindex=0}$, despite other states being less localized. In this study we fix $\Lparam=1$ to keep the state-dependent effect on localization length constant. When demonstrating strong localization in Fig.~\ref{fig:state_loc_dif} we show a single initial condition  $\ket{\pindex}$ with $\pindex \neq 0$ to avoid early delocalization. But in other figures the fidelity is averaged over all initial conditions to average out state-dependent effects. \newtext{This averaging of fidelity is motivated by the standard approach to observing the Lyapunov exponent in the fidelity decay rate \citep{benenti2002quantum}.}

{It is also worth noting that the initial conditions are chosen to be momentum eigenstates, which in the strongly localized limit are not perfectly localized since they are not quantum map eigenstates. Rather, in this limit the map eigenstates are complex superpositions of $\ket{\pindex}$ and $\ket{-\pindex}$.} As the quantum map causes the phase of each map eigenstate to evolve at a rate proportional to its quaisenergy, pairs of states will fall out of phase, causing an initial state $\ket{\pindex}$ to evolve to $\ket{-\pindex}$ after a phase difference $\pi$. However, in the localized limit the quasienergies of these map eigenstate pairs are close together, making this process long and irrelevant for the short time scales discussed in this paper. The fact that we reverse the map to calculate fidelity further reduces the relevance of this effect.

\begin{figure*}
	\centering
	\includegraphics[width=70mm]{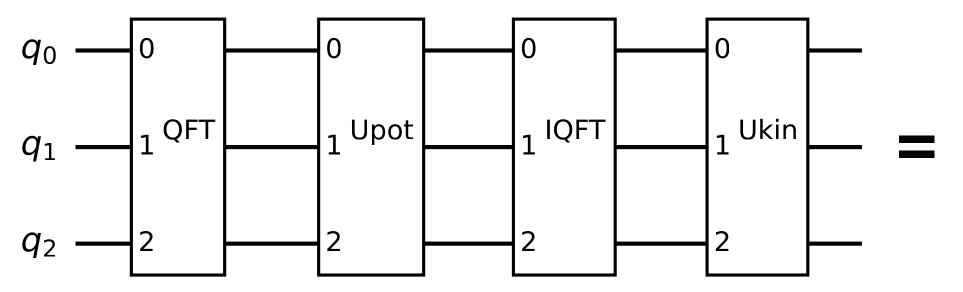}
	\includegraphics[width=130mm]{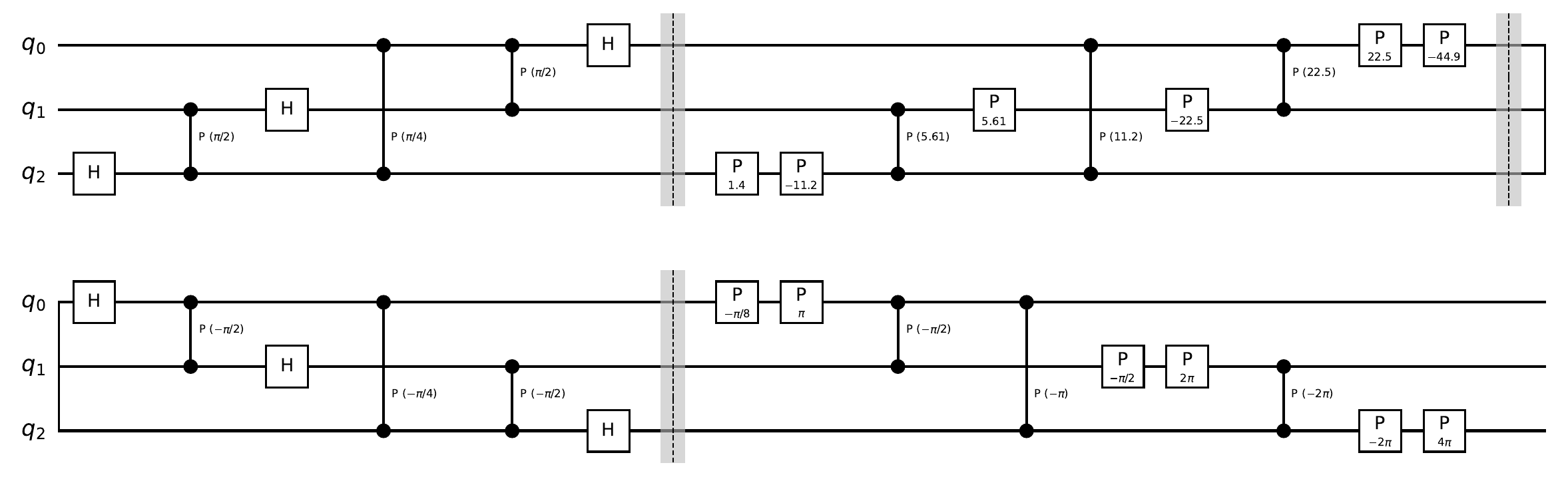}
	\caption{Circuit for a single forward map iteration of the three-qubit QSM algorithm from Eq.~\ref{eq:QSM_decomp_gates}, \newtext{both in block form and in algorithmic form} before conversion to hardware connectivity and transpilation to native gates. Two-qubit \code{CPHASE} gates are used. $\Upot$ and $\Ukin$ steps use \code{PHASE} and \code{CPHASE} gates, while $\Uqft$ steps use \code{CPHASE} and \code{H} gates. Using $\kparam=4.55$.}
	\label{fig:circuit_QSM}
\end{figure*}

\subsection{QSM algorithm} 
\label{sec:qsm_algorithm}

There is a natural mapping of the QSM to a qubit-based quantum computer. The $N$ momentum eigenstates can be mapped to the $2^{n}$ qubit states when $N=2^n$. The unitary $\Uqsm$ can then be implemented exactly in four steps \citep{porter2022observability,benenti2004quantum,georgeot2001exponential}, written compactly as
\begin{equation}
	\label{eq:QSM_decomp_simple}
	\Uqsm=\Ukin \Uqft^{-1} \Upot  \Uqft
\end{equation}
where the operators $\Ukin=\Uphase(\Tparam)$ and $\Upot=\Uphase(-k\beta^2)$ are many-qubit diagonal phase operators in the position and momentum bases respectively, defined in Eq.~\ref{eq:QSM_def}, and $\Uqft$ is the quantum Fourier transform used to alternate between the momentum and position bases. The diagonal phase operators can be implemented exactly due to a method for decomposing order-$P$ polynomial terms in the Hamiltonian into \newtext{polynomially many} $P$-qubit gates, given in Appendix \ref{appendix:gate_decomp} \citep{georgeot2001exponential}. A similar yet approximate algorithm exists for any quantum map whose Hamiltonian has separable kinetic and potential energy terms each with a convergent power series expansion, such as the standard map \citep{georgeot2001exponential} or kicked Harper model \citep{levi2004quantum}. An exact representation of trigonometric terms requires ancilla qubits. \newtext{Compared to these trigonometic potentials, the QSM algorithm has a particularly efficient algorithm due to its quadratic potential requiring only quadratically many two-qubit gates. While any polynomial length algorithm may be considered ``efficient", the QSM is the second-most efficient among quantum maps when using the Georgeot algorithm \citep{porter2022observability}.} For the $\Uqft$ steps we use the standard algorithm from IBM's library, which is exact, efficient, and requires no ancilla qubits \citep{nielsen2010quantum, IBM2021QFT}.

Using the derivation in Appendix \ref{appendix:gate_decomp}, the efficient circuit decomposition for the QSM is
\changedtext{
\begin{flalign}
	\label{eq:QSM_decomp_gates}
	\Uqft &= \prod_{j_1=0}^{n/2-1} \code{SWAP}_{j_1, n-1-j_1}
	\prod_{j_1=0}^{n-1} \left( \prod_{j_2>j_1}^{n-1} \code{CP}_{n-1-j_1, n-1-j_2}(\pi/2^{j_2-j_1}) \right) \code{H}_{n-1-j_1} \nonumber \\
	\Upot &= \prod_{j_1=0}^{n-1} \left( \prod_{ j_2>j_1}^{n-1} \code{CP}_{j_1, j_2}(\kparam \beta^2 2^{j_1 + j_2 }) \right) 
	\code{P}_{j_1}(\kparam \beta^2 2^{2j_1 -1} -\kparam \beta^2 N 2^{j_1-1})  \\
	\Ukin &= \prod_{j_1=0}^{n-1} \left( \prod_{ j_2>j_1}^{n-1} \code{CP}_{j_1, j_2}(-\Tparam 2^{j_1 + j_2 }) \right)
	\code{P}_{j_1}(-\Tparam 2^{2j_1 -1} +\Tparam N 2^{j_1-1}) \nonumber 
\end{flalign}
}
where \code{H} is a Hadamard gate, \code{P} and \code{CP} are one-qubit phase and two-qubit controlled-phase gates respectively{, and $\Uqft$ follows the Qiskit reverse-ordering convention}. \newtext{The second term in the argument of each \code{P} gate translates the domain to $p \in [-N/2, (N-1)/2]$, as described in Appendix~\ref{appendix:gate_decomp}.} This algorithm is shown for three qubits in Fig.~\ref{fig:circuit_QSM}, with the \code{SWAP} gates from $\Uqft$ having been eliminated by reversing the order of qubits during $\Upot$.

\newtext{Since our algorithm is exact while maintaining polynomial gate scaling, there is no clear benefit to methods that scale near-optimally with error such as quantum signal processing (QSP) \citep{low2017optimal}. Moreover, QSP has a large constant overhead cost and requires ancilla qubits which make it inappropriate for few-qubit applications. Another approach, a decomposition of diagonal unitaries to Walsh functions provided by \cite{welch2014efficient}, is also an approximation and so unnecessary here. Further, it is only efficient if the diagonal Hamiltonian is smooth enough that the number of Walsh functions $k$ is independent of the qubits $n$ at any fixed approximation error. However, as one reduces the error tolerance, the number of required gates would grow as $2^k$.}

\newtext{The initial conditions used in this work will vary between figures, as specified in their captions. However all basis states are valid initial conditions for exploring the dynamics of the QSM.}

\section{Experimental results}
\label{sec:main_results}

\subsection{Fidelity definition}
\label{sec:experiment_intro}

This section discusses the core experimental results. \changedtext{The main result concerns the Loschmidt echo fidelity, which is defined as the probability of evolving a state and then reversing that evolution perfectly in the presence of noise. In an experiment noise is naturally present in both the ``forward" and ``backward" steps.  For the QSM, an initial pure state $\ket{\psi}$ is evolved under the unitary $\Uqsm$ in the presence of noise for $t/2$ time steps, followed by its inverse $\Uqsm^{-1}$ with the same noise processes for $t/2$ time steps, resulting in the total noisy evolution $\Phi_{t}$ and the final mixed state $\matsigma(t) = \Phi_{t}(\ket{\psi} \bra{\psi})$. Then for an initial computational basis state $\ket{\psi} = \ket{p}$, the Loschmidt echo fidelity is
\begin{flalign}
	f(t) &= \bra{p} \matsigma(t) \ket{p} = \text{Prob}_t(\ket{p}).
\end{flalign}
For simulations and experiments we use $t \equiv 2*t_\text{fb}$ to define $t_\text{fb}$, the number of forward-and-back pairs of steps.  Hence, the fidelity is given by $f(t) \equiv f(2*t_\text{fb})$ and plotted with respect to $ t_\text{fb}$. See Sec.~\ref{sec:noise_models} for a more thorough description of the Loschmidt echo.
}

The main experimental result is that the rates of fidelity decay in  Fig.~\ref{fig:IBM_fidelity_decays} increase as the QSM increases its quantum kick parameter $\kparam$. This only alters the phases of transpiled \code{RZ} gates in $\Upot$ (see Appendix~\ref{appendix:optimize_IBM}), and does not change the \code{CNOT} gate count. Note, however, that the change in fidelity is correlated with the transition between localized and diffusive dynamics and saturates to limiting values at both low and high values of $\kparam$. \changedtext{This behavior resembles but is distinct from the semiclassical regime \citep{porter2022observability} where the fidelity decay rate can be controlled by the Lyapunov exponent, and, hence, the strength of chaos in the system. That regime it not yet experimentally accessible on IBM-Q.}

\newtext{The phrase ``diffusive dynamics" above refers to dynamics which, taken in the classical limit of infinite qubits, would recover the chaotic classical diffusion discussed in Sec.~\ref{sec:localization}. However in the context of small quantum systems, diffused quantum states fully explore the Hilbert space and have essentially random phases on each basis state. These states will be referred to as ``randomly entangled" states in Sec.~\ref{sec:lindblad}.}

\begin{table}
	\begin{center}
		\caption{Native gate counts and fidelity for executing each forward-and-back iteration of the quantum sawtooth map (QSM) experimentally on IBM-Q devices. \newtext{ \code{ibmq\_5\_yorktown} is included to compare the effect of higher connectivity.} To calculate forward-only gate counts as for Fig.~\ref{fig:IBM_loc_dif}, divide by two. (a) \code{CNOT} gate count when Qiskit transpiler attempts direct gate decomposition of the QSM unitary. (b) \code{CNOT} gate count when using the efficient algorithm Eq.~\ref{eq:QSM_decomp_gates} plus transpiler optimization on linear qubit connectivity. (c) Physical single-qubit gate count, not including virtual \code{RZ} gates. Range is over initial condition and dynamical map parameter $\kparam$. (d) Fidelity as measured by the one-step Loschmidt echo, partly from Fig.~\ref{fig:IBM_fidelity_decays}. Range is over $\kparam$, varied from diffusive to localizing dynamics, after averaging over initial conditions. Experiment on \code{ibmq\_5\_yorktown} performed on October 22, 2020 at 4:41pm EST{, experiments on \code{ibmq\_manila} performed on the date and time in Fig.~\ref{fig:IBM_fidelity_decays}}. Fidelities include measurement error, and at full decoherence reach $1/N$. 
			\label{table:gate_counts}}
		{\begin{tabular}{ p{24mm} p{20mm} p{26mm} p{29mm} p{24mm} } 
				\textbf{\, \newline \newline Device, \newline \# qubits \newline (connectivity) \newline \newline} & \textbf{\, \newline (a) \newline \# CNOTS \newline (raw Qiskit)} & \textbf{\, \newline (b) \newline \# CNOTs \newline (Eq.~\ref{eq:QSM_decomp_gates}+Qiskit)} & \textbf{\, \newline (c) \newline \# physical \newline single-qubit gates \newline (Eq.~\ref{eq:QSM_decomp_gates}+Qiskit)} & \textbf{\, \newline (d) \newline fidelity $f(1)$ \newline (Eq.~\ref{eq:QSM_decomp_gates}+Qiskit)} \\ 
				\code{ibmq\_manila}, n=2 & 4 & 4 & 12-16 & 0.88-0.90 \\
				\code{ibmq\_manila}, n=3 (linear) & 136 & 66 & 28-32 & 0.27-0.37 \\
				\code{ibmq\_5\_yorktown}, n=3 (triangular) \newline & 82 & 38 & 26-30 & 0.125 (1/N)\\
		\end{tabular}}
	\end{center}
\end{table}

\subsection{Dynamical localization}
\label{sec:experiment_loc_dif}

\begin{figure}
	\centering
	\includegraphics[width=80mm]{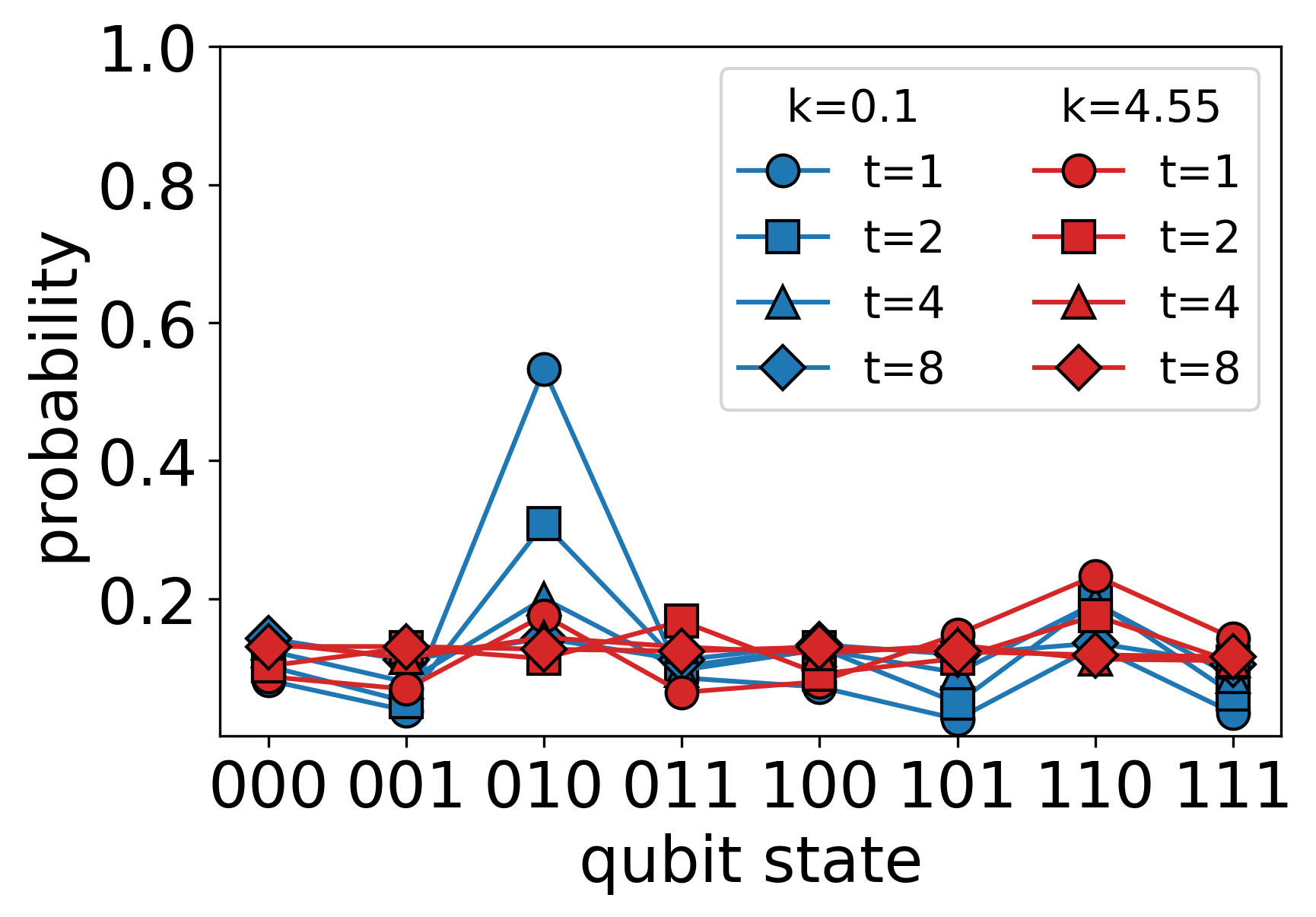}
	\caption{Dynamics of the QSM for three qubits on the \code{ibmq\_manila} device, showing localization ($\kparam=0.1$) and diffusion ($\kparam=4.55$) for $t=1,2,4,8$ and initial condition $\ket{p=-2}$ $(\ket{\psi}= \ket{010})$ for best localization. Compare to Fig.~\ref{fig:state_loc_dif}. Hereafter 8192 experimental shots were taken for each $\kparam$, $t$, and initial condition, giving statistical uncertainty $1/\sqrt{N_\text{shots}} \approx 1.1\%$. Experiment performed on April 11, 2022 at 10:19pm EST.  \code{ibmq\_manila} is re-calibrated every 1-2 hours to adjust for drift that can increase error.}
	\label{fig:IBM_loc_dif}
\end{figure}

One goal of simulating the QSM on present day hardware is to assess the hardware's ability to execute complex dynamical simulations. Fig.~\ref{fig:IBM_loc_dif} shows the results of simulating the QSM ``forward-only" {(ideally $\Uqsm$)} on the \code{ibmq\_manila} device in both the localized and diffusive regimes. The effect of dynamics is clearly apparent, and the localized state's probability is an informal metric of the fidelity of the quantum hardware. The raw data shows the localized state retains the maximum probability among the eight states through $t=7$. However the results of Fig.~\ref{fig:IBM_fidelity_decays} for simulating ``forward-and-back" {(ideally $\Uqsm^{-1} \Uqsm$)}  retain fidelity through at least $t_\text{fb}=5$, suggesting the forward-only localized case may retain information about its initial state through at least $t=10$.

The metric of the localized state's probability was used more thoroughly in Ref.~\citep{pizzamiglio2021dynamical}. Fig.~\ref{fig:IBM_loc_dif} does not easily compare to Ref.~\citep{pizzamiglio2021dynamical} because different map parameters were used, namely $\Lparam=1$ in this work and  $\Lparam=7$ in Ref.~\citep{pizzamiglio2021dynamical}. This significantly changes the eigenstates, independently of $\kparam$, and therefore changes the degree of localization. Different parameters were used in the present work to achieve more strongly localized dynamics, as explained in Sec.~\ref{sec:localization}.

\subsection{Fidelity and dynamics} 
\label{sec:experiment_fid}

\changedtext{The Loschmidt echo fidelity is a more quantitative metric of hardware ability for exploring the effect of dynamics on the fidelity. It is described in Sec.~\ref{sec:experiment_intro} and \ref{sec:types_noise_fidelity}.}

\begin{figure}
	\centering
	\includegraphics[width=80mm]{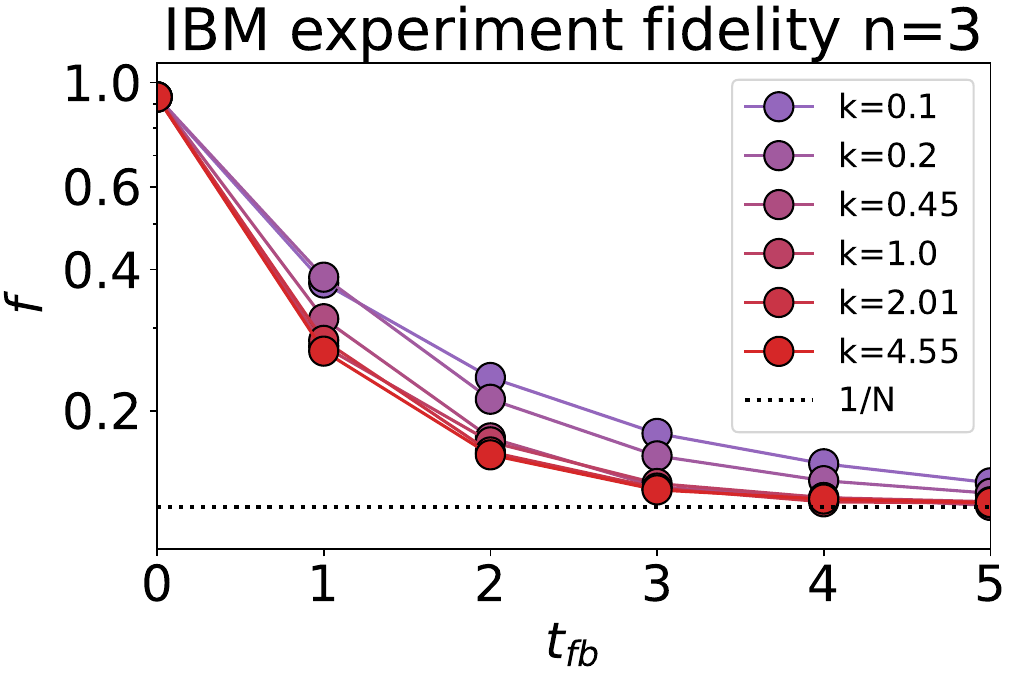}
	\caption{Average Loschmidt echo fidelity of the QSM on the \code{ibmq\_manila} device for  three qubits and varying $\kparam$. Localization occurs below $\kparam_\text{loc} \approx 1.87$. Data is averaged over all eight initial computational basis states. Statistical uncertainty per data point is $1/\sqrt{8192*8} \approx 0.4\%$. Number of CNOT gates per forward-and-back step is $M_\text{CNOT}=66$. The absolute fidelity gap at $t_\text{fb}=1$ between the most localized and most diffusive cases is 10.6\%. Experiment performed on January 28, 2022 at 2:44pm EST.}
	\label{fig:IBM_fidelity_decays}
\end{figure}

By varying the parameter $\kparam$ of the QSM the interaction between experimental noise and dynamics can be measured, as shown in Fig.~\ref{fig:IBM_fidelity_decays}. 
{As predicted by the \newtext{single-qubit} Lindblad noise models in Sec.~\ref{sec:lindblad} and Sec.~\ref{sec:lindblad_gate-based}, localized and diffusive dynamics have different fidelity decay rates during the various substeps of the simulation.} The experiment shows a continuous transition in the fidelity decay rate as the dynamics change from localized to diffusive. The noise models are used {in Sec.~\ref{sec:experiment_fit_lindblad}} to fit the data and extract effective decoherence parameters. 

Note that for $n=3, \Lparam=1$ as used here, the predicted transition to full diffusion should occur at $\kparam_\text{loc} \approx 1.87$. In the experiment, the largest three $\kparam$ values have indistinguishable fidelities up to a  $1.5\%$ absolute difference despite $\kparam=1.0$ being below the transition threshold. Further resolution of the observed transition value $\kparam_\text{loc}$ requires greater statistics and larger system size. The smallest three $\kparam$ values show a gradual transition from the strongly localized $\kparam=0.1$ to the weakly localized $\kparam=0.45$.


\begin{table}
	\begin{center}
		\caption{Comparison of IBM-Q's reported RB gate error to error extracted from a three-qubit experiment with localized ($\kparam=0.1$) or diffusive ($\kparam=4.55$) dynamics. Experimental error $\epsilon$ is calculated from fidelity decay $f(t)$ via $f(1) = (f(0)-1/2^n) (1-\epsilon)^{66} +1/2^n$. \label{table:ibm_cnot_comparison}}
		{\begin{tabular}{ p{45mm} p{36mm} p{35mm} }  
				\, \newline \, \newline & \textbf{\, \newline \code{CNOT} error (localized)}  & \textbf{\, \newline \code{CNOT} error (diffusive)} \\
				\code{ibmq\_manila} reported $n=2$ & 5.79E-3 & 5.79E-3 \\
				\code{ibmq\_manila} experiment $n=3$ & 1.76E-2 & 2.59E-2 \\
				ratio of experiment:reported \newline & 3.0 & 4.5 \\
		\end{tabular}}
	\end{center}
\end{table}

\subsection{Gate error}
\label{sec:experiment_gate_error}

The most direct metric for comparing our simulation results to reported metrics from IBM-Q is the \code{CNOT} gate error. This does not rely on any noise model, as it is a single-parameter fit (gate error) for each $\kparam$. In table \ref{table:ibm_cnot_comparison} error fits are reported, using just the first time step $f(1)$ and the SPAM error $f(0)$. The fidelity dependence on dynamics is codified here as gate error dependence on dynamics, with a factor of $1.5\times$ in gate error between the extreme dynamical cases.

More interestingly, this range of observed errors is $3.0\text{--}4.5\times$ worse than the error reported by IBM-Q on their online ``Systems" screen at the time the experiment was performed \citep{IBM2022systems}. Their reported error comes from standard two-qubit RB with a depolarizing noise model \citep{mckay2019three}. {The nature of RB circuits makes this difference in observed error unsurprising. However it is worth describing several possible contributors.}

{There are two main ways in which RB circuits reduce their observed error: the uniformly-randomized Clifford gates effectively depolarize the error, reducing the effect of coherent errors relative to general circuits; and the same randomization causes low-frequency noise to ``echo" and partially cancel, similar to a dynamical decoupling protocol. While these do simplify the interpretation of RB, they also make it a overly optimistic metric for predicting the performance of general circuits.}

{Another cause of the error difference is the extra crosstalk from adding a third qubit relative to two-qubit RB. The role of third qubit crosstalk has been investigated with simultaneous RB \citep{mckay2019three}, where average \code{CNOT} error per gate was found to increase from $1.63\mathrm{E}{-2}$ for two-qubit RB to $2.70\mathrm{E}{-2}$ for 2+1-qubit simultaneous RB, a factor increase due to crosstalk of $1.66$. How this factor varies across devices and between experiments is however less clear.}

{There are also differences between the dynamics of the QSM circuit and RB Clifford gates that may contribute.} Recent research suggests Clifford gates have different quantum scrambling properties than general unitary dynamics: out-of-time-order correlators (OTOCs), which are measures of quantum chaos and scrambling and close relatives of the Loschmidt echo fidelity, reach very different asymptotic values under Clifford and non-Clifford unitary evolution \citep{roberts2017chaos,leone2021isospectral}. This may relate to the Clifford group being a 2-design on qudits (and a 3-design on qubits) \citep{roberts2017chaos, webb2015clifford, zhu2017multiqubit}. Since OTOCs for quantum chaotic systems only grow exponentially until the Ehrenfest time $\tau_E$ \citep{hashimoto2017out}, the QSM which has $\tau_E \sim 1$ (see Appendix \ref{appendix:Lindblad_and_dynamics}) saturates OTOCs quickly. {While this faster, more thorough quantum scrambling may influence the fidelity, we leave the quantification of such an effect to future work.}

{Most of these effects could be captured in a process matrix picture, in which certain elements of the \code{CNOT} process matrix are more strongly enacted in the QSM circuit relative to an average over RB circuits. However, the effect of crosstalk goes beyond a static process matrix, as it causes the process matrix to depend on the number of qubits. This is because even spectator qubits can increase error \citep{mckay2019three}. The limited connectivity of IBM-Q devices becomes desirable here, as it reduces the number of neighbors per qubit which should limit the magnitude of crosstalk when scaling to many-qubit algorithms.}

Lastly, the depolarizing noise model determined through RB is clearly unable to capture or explain the fidelity dependence on dynamics. Its single parameter $\alpha_{2Q}$ of two-qubit gate error measures an average tendency towards the state $\matrho=I/N$, rather than capturing important details of how different density matrix elements contribute different rates of decay. Additionally, its focus on averaging over unitary errors, while mathematically convenient, is perhaps less appropriate than decoherence for describing present day superconducting quantum devices.

\section{Noise models} \label{sec:noise_models}

\subsection{Types of noise and fidelities}
\label{sec:types_noise_fidelity}

Present day quantum devices are impacted by many different types of noise. This motivates studies of the types of noise that are present and how the errors impact algorithms of interest. On the IBM-Q platform errors occur primarily during two-qubit gates which in the QSM algorithm contribute about $10\times$ more to the total error over single-qubit gates, as discussed in Appendix~\ref{appendix:optimize_IBM}. The types of error that occur may be incoherent Markovian relaxation and dephasing error ($T_1$ and $T_2$ respectively), coherent error, multi-qubit incoherent errors, or something else. \changedtext{Here three models of noise based on Lindblad master equations are considered for understanding the effects of errors: (1) an approximate analytic theory of  the effects of relaxation and dephasing, (2) a  Lindblad master equation simulation using \newtext{single-qubit Lindblad errors with} the four substeps of $\Uqsm$ in Eq.~\ref{eq:QSM_decomp_simple} rather than the actual gate decomposition, and (3) the IBM-Q Aer simulator which uses the exact gate decomposition and a Kraus noise process model based on \newtext{single-qubit} relaxation and dephasing. In Appendix~\ref{appendix:parameter_noise}, we prove that a stochastic Hamiltonian parametric noise model cannot explain the experimental results.}

The overall impact of noise can be studied by its effect on the rate of fidelity decay of our quantum system. When the noise is unitary, such as parametric noise (see Appendix~\ref{appendix:parameter_noise}), with total evolution $\matU_\epsilon$ and noise amplitude $\epsilon$, and a pure state $\ket{\psi}$ is used as an initial condition, the fidelity of the evolution can be measured by
\begin{equation}
	\changedtext{f(t)} = |\bra{\psi} \matU_{\epsilon'}^{-t} \matU_{\epsilon}^{t} \ket{\psi}|^2
\end{equation}
which is also known as the Loschmidt echo. $\epsilon'$ indicates the same magnitude as $\epsilon$ while being statistically independent. For non-unitary noise, such as Lindblad noise, one must use the more general density matrix formulation
\begin{flalign}
	f(\matrho, \matsigma) = \left( \Tr \sqrt{\sqrt{\matrho} \matsigma \sqrt{\matrho}} \right)^2
\end{flalign}
for ideal (initial) and noisy (final) density matrices $\matrho$ and $\matsigma$, respectively. {In the noiseless case $\matsigma$ should again return to its initial state $\matrho$ due to the forward-and-back evolution.} For an initial pure state $\matrho= \ketbra{\psi}$ this simplifies to \changedtext{
\begin{flalign}
	f(t) &= \bra{\psi} \matsigma(t) \ket{\psi} \nonumber \\
	\label{eq:fid_formula_as_elems}
	&= \sum_{i,j} \braket{\psi}{i} \bra{i} \matsigma(t) \ket{j} \braket{j}{\psi}  = \sum_{i,j} \sigma_{i,j}(t) \rho_{i,j}^* 
\end{flalign}
} \changedtext{which is used in Sec.s~\ref{sec:experiment_intro} and \ref{sec:lindblad}.}

Contrary to previous studies, noise here occurs during both forward and backward evolution, to connect simulation to experiment. \changedtext{This increases the total time relative to number of forward map steps by a factor of two, and therefore the observed fidelity decay rate by the same factor.} In all cases we average the fidelity over the $N$ initial conditions $\ket{\pindex}$ of the computational basis in order to study the average dynamics \citep{benenti2002quantum}.

\subsection{Effects of dynamics on Lindblad noise}
\label{sec:lindblad}

The Lindblad master equation is the most general type of completely positive trace-preserving (CPTP) Markovian master equation \citep{manzano2020short, pearle2012simple, lindblad1976generators, gorini1976completely, stephane2006open, gardiner2000quantum}. It can be written {in dimensionless form} as \changedtext{
\begin{flalign}
	\label{eq:lindblad_eq}
	\frac{\partial \matsigma}{\partial t} = -i [\matH,\matsigma] + \sum_i \nu_i \left(\matL_i \matsigma \matL_i^\dagger - \frac{1}{2} \{ \matL_i^\dagger \matL_i, \matsigma \} \right)
\end{flalign}
}for general Lindblad operators $\{\matL_i\}$, where $t = t_\text{phys}/T_\text{step}$ has units in the number of map steps{, $T_\text{step}$ is the dimensioned time to execute $\Uqsm$ on hardware, $\matH=\matH_\text{phys} T_\text{step}/\hbar$ with the forms of $\matH$ and $\matH_\text{phys}$ depending on the substep of the algorithm}, and $\nu_i = \nu_{i, \text{phys}} T_\text{step}$. \changedtext{The Loschmidt echo fidelity $f(t)$ comes from reversing the unitary evolution from $\matH$ without inverting the noise processes $\matL_i$.}

This is modified in the case of parametric noise, see Appendix~\ref{appendix:parameter_noise} for details.

As for the Lindblad operators in Eq.~\ref{eq:lindblad_eq}, we use $2n$ single-qubit operators $\{\matL_{1,j}, \matL_{2,j}\}$, defined as
\begin{flalign}
	\label{lindblad_L1}
	\matL_{1,j} &= I \otimes \dots
	\begin{pmatrix}
		0 & 1 \\
		0 & 0
	\end{pmatrix}_j
	\dots \otimes I \\
	\label{lindblad_L2}
	\matL_{2,j} &=  I \otimes \dots
	\begin{pmatrix}
		0 & 0 \\
		0 & 1
	\end{pmatrix}_j
	\dots \otimes I,
\end{flalign}
causing relaxation to the ground state at rate $\nu_1$ and pure dephasing at rate $\nu_2$ for each qubit $j$, as described in Appendix \ref{appendix:Lindblad_and_dynamics}. These relate to relaxation time $T_1$ and total dephasing time $T_2$ via
\begin{eqnarray}
	\label{eq:Ts_and_nus}
	1/T_1 = \nu_1, \qquad 1/T_2 = \nu_1/2 + \nu_2/2
\end{eqnarray}
as can be seen by comparing Appendix \ref{appendix:Lindblad_and_dynamics} to Ref.~\citep{krantz2019quantum}. For calculating the fidelity, the ideal expected density matrix, which is also the initial density matrix, will be denoted $\matrho$, and the noisy evolving density matrix will be denoted $\matsigma$.

The main interaction between dynamics and Lindblad noise is that different types of dynamics have different typical density matrices that are influenced by each decoherence effect to different degrees. These include on-diagonal relaxation (of computational basis states), off-diagonal relaxation (of superpositions of basis states), and off-diagonal dephasing. In Appendix \ref{appendix:Lindblad_and_dynamics}, we analytically determine the decay of fidelity for pure states in the two dynamical limits of being highly localized and highly diffusive, as well as the case of uniform superposition, all three of which are summarized in this section. Briefly, the fidelity of localized pure states is affected only by on-diagonal relaxation due to their lack of off-diagonal terms, while the fidelity of diffusive pure states is affected only by off-diagonal relaxation and dephasing due to a cancellation of on-diagonal effects. The key findings here are that the $\nu_1$ dependence is the same for both despite differing origins while the $\nu_2$ dependence is greater in the diffusive case.

Starting with the fully localized case which has $\kparam=0$, initial conditions $\ket{\psi}=\ket{\pindex}$ of computational basis states, and fidelity $f=\sigma_{p,p}$, the evolution of $\ket{\psi}$ due to the diagonal Hamiltonian has no effect on the density matrix $\matsigma$ and therefore on the fidelity. The Lindblad evolution acts alone, causing relaxation of each qubit towards the ground state at rate $\nu_1$. If one averages the fidelity over all possible initial states $\ket{\pindex}$, the result is
\begin{flalign}
	\label{eq:fid_unentangled}
	f_L(t,n) =& \left( 1 + e^{-2 \nu_\text{single} t} \right)^n / 2^n
\end{flalign}
so that the initial effective decay rate is
\begin{flalign}
	\label{eq:fid_1st_order} 
	f_L(t,n) &\approx 1 - \nu_\text{eff} t + O(\nu_\text{eff}^2 t^2) \approx e^{-\nu_\text{eff} t } \\
	\nu_\text{eff}(n) &= n \nu_\text{single} \nonumber
\end{flalign}
where 
\begin{eqnarray}
	\label{eq:rate_localized}
	\nu_\text{single} &=& \nu_1/2 = 1/2 T_1
\end{eqnarray}
for the localized case. The factor of $1/2$ derives from the average $1/2$ chance of each qubit starting in the excited state. The factor of $n$ due to $n$ qubits decaying is an important aspect of interpreting measured $T_1$ and $T_2$ times. Dynamics that are less than fully localized will start to show effects of the diffusive case described below.

Between the localized and diffusive cases, another interesting case is the decay of a uniform superposition state, where all $n$ qubits are in $\ket{+}$ states which are unentangled relative to the computational basis. The fidelity is a product of the single-qubit fidelities. This may look like a simple model of a diffused state since the probability has spread to all states equally, but a general diffused state would also be entangled with random, independent phases on each of the $2^n$ states. The uniform superposition state is then a useful test of the relative roles of spreading probability versus entanglement in quantum state diffusion. Only half of the single-qubit fidelity decays, similar to the averaging factor of one half in the localized case. The same Eq.~\ref{eq:fid_unentangled} and Eq.~\ref{eq:fid_1st_order} apply with
\begin{eqnarray}
	\label{eq:rate_superposition}
	\nu_\text{single} &=& (\nu_1+\nu_2)/4 = 1/2 T_2.
\end{eqnarray}

In the diffusive case the Hamiltonian evolution becomes important, but for the QSM we have found it can be abstracted away to simplify the problem and still produce predictions that agree well with simulation\newtext{, as we will describe here}. To simplify, note that chaotic mixing occurs on faster time scales than Lindblad decay. Then the density matrix can be expected to quickly reach a randomly entangled pure state\newtext{, having (approximately) uniformly spread probability and random relative phases,} after which the Hamiltonian evolution has little qualitative effect aside from rapidly changing the precise values of the  phases. \newtext{(See Eq.~\ref{eq:apx_entangled_pure_state} for a model of a randomly entangled state.)} Fidelity decay during diffusive dynamics can then be approximated as the decay of this random pure state under Lindblad evolution. \newtext{Averaging over initial conditions (basis states) allows this random state to be analyzed via the average behavior of the statistical ensemble it belongs to. For a chaotic system, this corresponds to averaging over the entire accessible phase space. Such averaging} contributes to an averaging over these random phases, causing the phase-dependent terms in $f$ to vanish. Inspecting $f=\sum_{i,j} \sigma_{i,j} \rho_{i,j}^*$ shows that only $\sigma_{i,j}$ terms that retain their initial phase from $\rho_{i,j}$ survive due to phase cancellation in $f$. Surprisingly, this causes the qubits to effectively disentangle for off-diagonal terms $\sigma_{i,j} \rho_{i,j}^*$, in a manner describable by two pieces: a non-trace-preserving single-qubit decay and a trace-preserving correction. Along the $n$-qubit diagonal the flat $\rho_{i,i}=1/2^n$ causes relaxation effects across $\sigma_{i,i} \rho_{i,i}^*$ to cancel out. This leaves only effects from dephasing and off-diagonal relaxation, where the random phase averaging removes the gain to lower states but not the loss from higher states. The exact fidelity expression for the diffusive case is provided in Eq.~\ref{eq:apx_fid_random_phase_simplified} in Appendix \ref{appendix:Lindblad_and_dynamics}, but its initial decay rate using Eq.~\ref{eq:fid_1st_order} is simply
\begin{eqnarray}
	\label{eq:rate_diffusive}
	\nu_\text{single} &=& \nu_1/2+ \nu_2/4 \\
	&=& 1/4 T_1 + 1/2 T_2 \geq 1/2 T_1, \nonumber
\end{eqnarray}
which is strictly faster than both the localized and superposition cases, and any unentangled case, when  $\nu_1, \nu_2 > 0$. Random entanglement has the same dephasing rate as unentangled superposition, but it has an additional relaxation effect due to the average over random phases between the excited and decaying states.

\begin{figure}
	\centering
	\includegraphics[width=60mm]{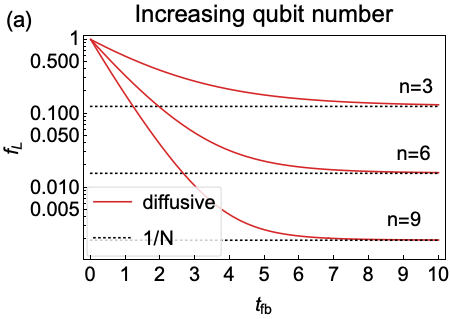}
	\includegraphics[width=60mm]{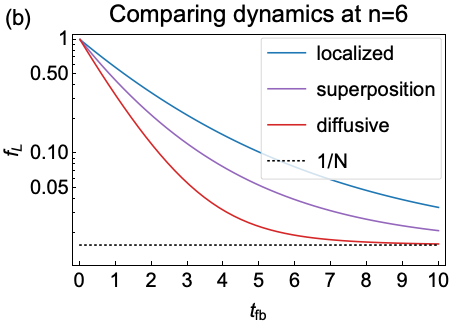}
	\caption{Theoretical Lindblad fidelity evolution {from forward-and-back noise} for $\nu_1=0.1$ and $\nu_2=0.2$: (a) comparing the decay of a diffusive state for three, six, and nine qubits; (b) comparing the decay of localized, superposition, and diffusive states for six qubits. The average fidelity plateaus at the uniformly mixed limit $1/N$ when no information about the original state remains.}
	\label{fig:theory_demos}
\end{figure}

Figure \ref{fig:theory_demos} compares the full analytic fidelity evolution for each dynamical case. Each fidelity has the expected feature of starting with an exponential decay that gradually relaxes to the uniformly mixed value of $1/N$. In superconducting qubits the $\nu_2$ rate often dominates, so three types of dynamics are compared for that case in Fig.~\ref{fig:theory_demos}(b). These fidelity decay rates do not account for the gate implementation used in experiment, which is partly rectified in the next section.

\subsection{Gate-based Lindblad model}
\label{sec:lindblad_gate-based}

The theoretical expressions for the localized and diffusive dynamical cases can now be modified to a gate-based form to better match experimental observations that the majority of errors occur during certain specific gates. Since two-qubit gates are the dominant source of error on IBM-Q, one can model circuit error to lowest order as solely being due to the two-qubit gates causing an enhanced Lindblad decay for each of the target qubits. {Further rationale for this model is provided in Appendix \ref{appendix:rationale}.} Averaged over the possible two-qubit subsystems on which the gates act during a circuit, this is approximately described by the dynamics-dependent expressions previously derived, with number of qubits set to two. If the gates are performed serially then the serial gate-based fidelity $f_\text{GB, S}(t,n)$ can be given in terms of the dynamics-dependent expressions for the Lindblad fidelity $f_\text{L}(t',n')$ \newtext{of Eq.~\ref{eq:fid_1st_order}} as
\begin{flalign}
	\label{eq:fid_gb_s}
	f_\text{GB,S}(t,n) = (f_\text{L}(1/M, 2))^{M t} (1-1/2^n) + 1/2^n
\end{flalign}
where $M$ is the number of gates per map step, $t=1$ is the time to complete a map step, $n$ is the total number of qubits in the circuit, $n'=2$ is the number of decaying qubits per gate duration, and $t'=1/M$ is the time per gate duration as a fraction of a map step. Note the late-time (average) fidelity should always approach $1/2^n=1/N$. For large $M$ the Lindblad expression simplifies to $f_\text{L}(t \ll 1,n) \approx e^{-n \nu_\text{single} t}$, so 
\begin{flalign}
	\label{eq:fid_gb_s_approx}
	f_\text{GB,S}(t,n) \approx e^{-2 \nu_\text{single} t} (1-1/2^n) + 1/2^n.
\end{flalign}
This model will be seen in Sec.~\ref{sec:experiment_fit_lindblad} to qualitatively match experimental results for $n=3$. Compared to a model that assumes all three qubits are decaying at equal rates, when fit to experimental data this gate-based model measures $\nu_1$ and $\nu_2$ that are about $3/2$ larger.

Performing some gates in parallel would increase the average number of simultaneously targeted qubits from two to $n_\text{eff} \equiv 2*\lfloor n/2 \rfloor \leq n$ and decrease gate depth from $M$ to $D=2M/n_\text{eff}$. Then the fidelity would be
\begin{eqnarray}
	\label{eq:fid_gb_p}
	f_\text{GB,P}(t,n) = (f_\text{L}(1/D, n_\text{eff}))^{D t}  (1-1/2^n) + 1/2^n.
\end{eqnarray}
For large $M$ this simplifies to 
\begin{eqnarray}
	\label{eq:fid_gb_p_approx}
	f_\text{GB,P}(t,n) \approx e^{-n_\text{eff} \nu_\text{single} t} (1-1/2^n) + 1/2^n.
\end{eqnarray}
Since $\nu_\text{single} = \nu_\text{single,phys} T_\text{step}$ depends on the physical map step time $T_\text{step}$, which is $n_\text{eff}/2$ times shorter for the parallel case compared to serial, the two cases yield the same fidelity as a function of map step $t$. This just reflects the fact that total fidelity is the product of gate fidelities when gate error is much greater than idle error. {As gate error approaches idle error and the number of idle qubits increases, idle errors complicate this analysis and a benefit to parallelization appears.}

\begin{figure}
	\centering
	\includegraphics[width=60mm]{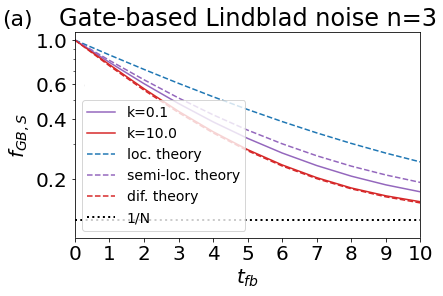}
	\includegraphics[width=60mm]{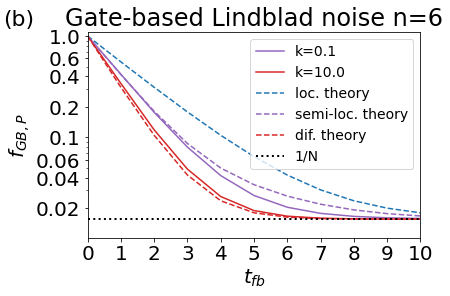}
	\caption{Comparing Lindblad evolutions from full simulation (solid) to the theoretical steady state approximation (dashed) for the fully localized (blue), semi-localized (purple, $\kparam=0.1$) and fully diffusive (red, $\kparam=10.0$) cases. Using $\nu_1=0.1$ and $\nu_2=0.2$. (a) $n=3$ ($n_\text{eff}=2$) and (b) $n=6$ ($n_\text{eff}=6$).}
	\label{fig:theory_vs_sim}
\end{figure}

To test the theoretical predictions of Eq.~\ref{eq:fid_gb_s} and \ref{eq:fid_gb_p}, we use QuTiP's \citep{johansson2012qutip} master equation solver. To partially mimic a circuit model we simulate the four unitary steps of Eq.~\ref{eq:QSM_decomp_simple} sequentially, rather than simulating the whole Hamiltonian at once. This turns out to be especially important for the localized case, as described below. For the three-qubit simulation we also mimic two-qubit gates by only applying Lindblad operators to two qubits at a time, alternating the qubit pairs [0,1] and [1,2] to mimic the near-linear connectivity on many IBM-Q devices. This matches the effective two-qubit decay rate from theory.

The simulation results in Fig.~\ref{fig:theory_vs_sim} require a modified interpretation of the theory, as shown  for three qubits with serial gates and six qubits with parallel gates. In the localized case, a straightforward Hamiltonian simulation would keep the state localized during the whole map step and would fit well to the theoretical localized decay. But decomposing the evolution to substeps (or further to gates) reveals that the quantum Fourier transforms (QFTs) take the localized state to and from a \newtext{delocalized state} during half of each map step. \newtext{This delocalized state has phases that are even spaced around the unit circle and thus will usually average to zero. This effect is similar to the derivation of the diffusive fidelity decay rate in Appendix~\ref{appendix:Lindblad_and_dynamics}, suggesting that the diffusive decay rate could be appropriate here. This is empirically supported by the simulation results in Fig.~\ref{fig:theory_vs_sim} for $\kparam=0.1$, which fit better to a half-localized, half-diffusive model as shown than to a half-localized, half-superposition model which would decrease $\nu_\text{single}$ by $\nu_1/8$. Therefore we model the map step fidelity as $f_\text{semi-loc} \approx f_\text{loc}(t/2) f_\text{dif}(t/2) \approx (f_\text{loc} f_\text{dif})^{1/2}$. {Using Eq.~\ref{eq:fid_1st_order} with Eq.~\ref{eq:rate_localized} for $f_{loc}$ and Eq.~\ref{eq:rate_diffusive} for $f_{dif}$} implies a ``semi-localized" decay rate given by}
\begin{flalign}
	\label{eq:rate_semi_localized}
	\nu_\text{single} = \nu_1/2 + \nu_2/8.
\end{flalign}
With this modified theoretical prediction, the difference between localized simulation and semi-localized theory is small. One cause of the remaining difference in Fig.~\ref{fig:theory_vs_sim} is the small non-zero $\kparam$ allows small, random amplitudes on other states. This creates some random entanglement that enhances the decay rate during the localized part of the algorithm. This explains some but not all of the difference, with the unexplained portion growing with increasing $n$ and $t_\text{fb}$. For example, at $n=3$ the effect of non-zero $\kparam$ accounts for $\ge 50\%$ of the difference for $t_\text{fb} \geq 5$, but at $n=6$ it explains only $12\%$ of the difference at $t_\text{fb} = 5$. {(Simulations of this not shown.) The remaining difference between theory and numerics is attributable to equating between a randomly diffused state and the QFT of a localized state in deriving  Eq.~\ref{eq:rate_semi_localized}.}

In the diffusive case the difference between theory and numerics is generally smaller despite neglect of the exact Hamiltonian evolution. This suggests that the effect of the precise Hamiltonian evolution on the fidelity is subdominant.

While this model only includes error from two-qubit gates, it could be modified to include other gates. The largest correction would come from idle gates for idling qubits, which have their own decay rates $\nu_1$ and $\nu_2$. Idle gates are important because \code{CNOT} gate error comes primarily from the duration of the gate, implying that any qubit idling in parallel with a \code{CNOT} is experiencing a significant fraction of the error. For more details see Appendix~\ref{appendix:circuit_scaling} for gate specifics and table \ref{table:ibm_fit_comparison} for experimental fits.

\subsection{IBM Aer gate-based Lindblad simulator}
\label{sec:ibm_sim_noise}

IBM offers several options for artificial noisy simulators. One option is their device-specific noise models \citep{IBM2021noise} which are calibrated to each device's reported errors. Reported $T_1$ and $T_2$ times and per-gate error rates from randomized benchmarking are used as input for a combined relaxation, dephasing, and depolarization error model. However, because these models are not customizable, they cannot be fit to experimental data.

\begin{figure}
	\centering
	\includegraphics[width=80mm]{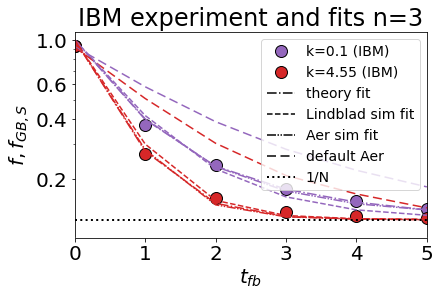}
	\caption{Numerical fits of several models to the data in Fig.~\ref{fig:IBM_fidelity_decays} for the extreme conditions of localized (purple, $\kparam=0.1$) and diffusive (red, $\kparam=4.55$) dynamics. Models described in the main text. \newtext{Note for both values of $k$ the plots for the theory fit and Aer sim fit overlap and resemble a solid line.}}
	\label{fig:IBM_fit_sims}
\end{figure}

We instead use a custom IBM Aer noise model. Closest to our gate-based Linblad model is a simple Aer model based on the function \code{thermal\_relaxation\_error} with parameters $T_1$ and $T_2$ and an effective temperature that we set to zero (the default option) \citep{IBM2021source}. This function applies a single-qubit Lindblad decay to each qubit targeted by a gate based on the duration of the gate, after the gate is done. When $T_2>T_1$ it applies Kraus operators, but when $T_2<T_1$ as is common it probabilistically applies single-qubit gates instead. We modify this function to always apply Kraus operators. This model will be used in Fig.~\ref{fig:IBM_fit_sims} to compare the relative importance of specifying the full gate dynamics as in Aer versus simplifying to the four unitary substeps Eq.~\ref{eq:QSM_decomp_simple} as in the gate-based Lindblad model.

\begin{table}
	\begin{center}
		\caption{Parameter fits from Fig.~\ref{fig:IBM_fit_sims}. All values are averaged over three qubits connected in a line on \code{ibmq\_manila}. $T_\text{1,phys}$ and $T_\text{2,phys}$ values are converted from simulations using $T_\text{1,phys} = T_\text{step} / \nu_1$ and $T_\text{2,phys} = T_\text{step}*2 / (\nu_1+\nu_2)$ where $T_\text{step}=33*350 ns$. {Errors are propagated by assuming zero covariance of $\nu_1$ and $\nu_2$, as the Lindblad model stipulates.} Relaxation $\nu_1$ and pure dephasing $\nu_2$ are dimensionless decay rates per single-direction map step forwards or backwards in time.  The analytic theory applies continuous Lindblad decay for two qubits for each of $66$ gates per map step. The Lindblad simulation continuously applies Lindblad operators to two qubits in alternating pairs during the $8$ algorithm substeps. The modified Aer simulator applies Lindblad decay to one and two qubits as Kraus operators after each gate. \label{table:ibm_fit_comparison}}
		{\begin{tabular}{ p{45mm} p{19mm} p{19mm} p{20mm} p{20mm} }  
				\, \newline \, \newline & \, \newline {\textbf{$\bm{T_{1,\text{phys}} (\mu s)}$}}& \, \newline \textbf{$\bm{T_{2,\text{phys}} (\mu s)}$} & \, \newline $\bm{\nu_1}$ & \, \newline $\bm{\nu_2}$ \\
				\code{ibmq\_manila} reported $n=1$ & 143 & 37.4 & 0.081 & 0.537 \\
				theory fit to experiment $n=3$ & 34.6 $\pm$ 1.7 & 14.4 $\pm$ 0.6 & 0.334 $\pm$ 0.016 & 1.271 $\pm$ 0.068 \\
				Lindblad simulation fit $n=3$ & 90.1 $\pm$ 41.6 & 14.3 $\pm$ 1.7 & 0.128 $\pm$ 0.059 & 1.486 $\pm$ 0.185 \\
				Aer simulator fit $n=3$ \newline & 250 $\pm$ 104 & 13.4 $\pm$ 0.8 & 0.046 $\pm$ 0.019 & 1.68 $\pm$ 0.10 \\
		\end{tabular}}
	\end{center}
\end{table}

\subsection{Fitting theory to experiment}
\label{sec:experiment_fit_lindblad}

{Having now described our noise models and performed derivations and simulations that qualitatively agree with experiment, it is interesting to fit those models to experiment to extract phenomenological decoherence parameters. The computational models can be fit through standard variational methods. The analytic gate-based Lindblad model from Eq.~\ref{eq:fid_gb_s_approx} has a fidelity that decays as \changedtext{
\begin{flalign}
	f_\text{GB,S}(2 t_\text{fb} ) & \approx e^{-n_\text{eff} \nu_\text{single} * 2 t_\text{fb} } (1-1/2^n) + 1/2^n
\end{flalign}
}where $n_\text{eff}=2$ for the two active qubits per gate, \changedtext{$t \equiv 2 t_\text{fb}$ with $t_\text{fb}$ as the number of forward-and-back pairs of map steps}, and $\nu_\text{single}$ depends on the dynamics of the map. \changedtext{After adjusting for the dynamics of the full QSM algorithm, the effect of localized dynamics is that}
\begin{flalign}
	\nu_\text{single} = \nu_1/2 + \nu_2/8 = 3/8 T_1 + 1/4 T_2
\end{flalign}
while the effect of diffusive dynamics is that
\begin{flalign}
	\nu_\text{single} &= \nu_1/2+ \nu_2/4 = 1/4 T_1 + 1/2 T_2
\end{flalign}
as explained in Sec.~\ref{sec:lindblad} and \ref{sec:lindblad_gate-based}, \newtext{with further details in Appendix~\ref{appendix:Lindblad_and_dynamics}.}

To compare the experiment of Fig.~\ref{fig:IBM_fidelity_decays} to {this theory}, it is convenient to focus on the extreme cases of $\kparam=0.1$ and $\kparam=4.55$. An account of continuously changing $\kparam$ in a Lindblad model and its effect on fidelity would require a careful application of localization length to the Lindblad model, which is beyond the scope of this paper. 

Fig.~\ref{fig:IBM_fit_sims} shows numerical fits to experimental fidelity of extreme $\kparam$ for the gate-based Lindblad theory, the gate-based Lindblad simulations, and an IBM-Q Aer simulator model, all described in Sec.~\ref{sec:noise_models}. Also included is the default Aer model from the function \code{AerSimulator.from\_backend(FakeManila())}, which is designed to fit the error measured by RB. 

The default Aer model underestimates error, but it only underestimates it by a factor of $1.5$ instead of $3.0\text{--}4.5$ from table \ref{table:ibm_cnot_comparison}. This is because it draws from a previous device calibration when the relevant gates were $2.0\text{--}3.0\times$ worse, artificially improving its accuracy. On average it is likely to be less accurate for the three-qubit QSM than shown here.

{The three higher accuracy model fits demonstrate different approaches.} The Lindblad theory fit uses the gate-based model Eq.~\ref{eq:fid_gb_s_approx} with rates Eq.~\ref{eq:rate_semi_localized} for localized dynamics and Eq.~\ref{eq:rate_diffusive} for diffusive dynamics. The Lindblad simulation fit applies single-qubit  relaxation and dephasing of uniform rates $\nu_1$ and $\nu_2$ continuously to two out of three qubits, alternating between qubits 0 and 1 or 1 and 2. This is applied during the four unitary steps per Eq.~\ref{eq:QSM_decomp_simple} evolved forward-and-back with QuTiP's \code{mesolve} function. The Aer simulator fit applies two-qubit and single-qubit gates followed by Kraus operators on the targeted qubits that are determined from decoherence parameters $T_{1,\text{phys}}$ and $T_{2,\text{phys}}$.

These models of fidelity all have just two parameters: $\nu_1$ for relaxation and $\nu_2$ for dephasing, which relate to the physical decoherence times through Eq.~\ref{eq:Ts_and_nus} and $T_1 = T_{1,\text{phys}}/T_\text{step}, T_2 = T_{2,\text{phys}}/T_\text{step}$. $T_\text{step}$ is the time to complete a single forward or backward map step on hardware, with $T_\text{step}=33*350 ns$ to match $33$ \code{CNOT} gates of average duration $\approx 350 ns$ on each of the forward and backward simulation steps. \newtext{(This was an accurate average gate time when these experiments were performed, though since then IBM's two-qubit gates have changed in type and duration. \citep{IBM2022systems}.)} This only considers time spent in \code{CNOT} gates, in accordance with the single-gate-based Lindblad model. The Aer simulator model uses the same $T_\text{step}$ to convert between parameters, despite including single-qubit gates decaying at the same rate as the \code{CNOT} gates, since the single-qubit gates contribute $<5\%$ to the total duration of gates {($1/10$ the duration of a \code{CNOT} and $< 1/2$ as many gates)}. Since it directly fits $T_{1,\text{phys}}$ and $T_{2,\text{phys}}$, this extra error is in the converted $\nu_1$ and $\nu_2$ values. Measurement error is neglected in all models, which requires additional parameters to describe the probability of each state being misclassified, and which has small effect on the large exponential decay rates measured here.

{The fit of all three models in Fig.~\ref{fig:IBM_fit_sims} is qualitatively quite good. The Lindblad simulation is a slightly worse fit than the others, which is surprising since it should be more accurate than the corresponding simplified theoretical model. The late time fit may appear to have higher error, but this is only an illusion of plotting log-fidelity which accentuates the errors when the fidelity is small.}

Table~\ref{table:ibm_fit_comparison} provides the quantitative parameter fits. The most striking result is the consistent values of $T_2$  across models, as it is also the dominant error source. {$T_2$ measures the dephasing of states that are in superposition, and the QSM with diffusive dynamics produces randomly entangled states that dephase similarly to superposition states (see Appendix \ref{appendix:Lindblad_and_dynamics}), so it is well-suited to benchmark an effective version of this noise process.} {The agreement between models suggests the analytical model might be usable for lightweight extraction of the effective circuit $T_2$ time.}

{The variance across the fit $\nu_1$ and $\nu_2$ values suggests differing effective weights on each across the three models. The analytic model suggests that $\nu_2$ is the sole cause of a difference between localized and diffusive fidelity decay rates. However three qubits is a small system and has poor random phase averaging for the diffusive case, which could alter the relative roles of $\nu_1$ and $\nu_2$. For larger system sizes the simplified theoretical model should have better agreement with the simulation results.}

The average $T_2$ fit of the three models is $14.0 \mu s$, which is $2.7\times$ smaller than the reported single-qubit idle $T_2$ time from IBM-Q. {It is no surprise that a complex circuit decoheres more quickly than idle qubits.} The experimental $T_1$ fit values are less consistent, though this is due mostly to the small values of $\nu_1$ being close to zero and therefore being similar in magnitude to the error bars, making them difficult to resolve. In particular the observed $T_1$ from the Aer simulator fit is larger than the reported idle value, casting some doubt on the model's accuracy with respect to $T_1$.

{The ratio of idle $T_2$ to circuit \code{CNOT} $T_2$ is also a useful test of the question posed in Appendix~\ref{appendix:circuit_scaling}: How does the error of a \code{CNOT} gate $\epsilon_\text{CNOT}$ compare to the error of an idle gate of the same duration $\epsilon_\text{IDLE}$? This test is hampered somewhat by IBM-Q reporting a $T_2$ value measured by the Hahn echo $T_{2E}$. That experiment is less sensitive to ``inhomogeneous broadening", that is low-frequency fluctuations that can cause large differences from trial to trial \citep{krantz2019quantum}. {Since $T_{2E} \ge T_2$, this only provides an upper bound, $\epsilon_\text{CNOT} / \epsilon_\text{IDLE} \lesssim T_{2E,\text{IDLE}} / T_{2,\text{CNOT}} \approx 3$, when the effective $T_1$ is negligible. This bound is right in between the limits of $\epsilon_\text{CNOT} = \epsilon_\text{IDLE}$ and $\epsilon_\text{CNOT} \gg \epsilon_\text{IDLE}$, suggesting the benefit of gate parallelization must be treated carefully. In particular, the approximation of negligible idle error in Sec.~\ref{sec:lindblad_gate-based}, which was reasonable to first order here, will become poor for serially executed circuits if even a few more idle qubits are added.}}

\section{Conclusion} \label{sec:conclusion}

{We performed digital quantum simulations of the quantum sawtooth map (QSM), a toy model of wave-particle interactions in plasmas, in order to understand the interaction between chaotic dynamics and noise on quantum hardware platforms.} 
We chose to simulate the quantum sawtooth map, a quantized analog of the {classical sawtooth map}, because it is one of the simplest possible such maps with one of the shortest possible gate sequences \newtext{and could be used to test the Lyapunov algorithm on future NISQ devices}. 
The resulting dynamics can be tuned from integrable and localized to chaotic and diffusive and leverages the noise that is naturally present on NISQ quantum hardware platforms to {benchmark} quantum computations that are relevant to particle transport in plasmas.
An important conclusion of our work is that the dynamics of a simulation can strongly influence the fidelity decay rate by changing the relative impact of different noise processes on the fidelity. 

{In this study the important relationship between dynamics and noise in a quantum simulation and the resulting effect on fidelity were examined in a minimal and experimentally accessible context. A gate- and qubit-efficient digital quantum simulation of the quantum sawtooth map was performed on the open-access IBM-Q platform, and then simulated with a gate-based Lindblad noise model which was shown to qualitatively capture the intended effect. Experiment, numerics, and analytic theory all demonstrated that as this quantum map is varied from near-integrable localized dynamics to diffusive and randomly entangled dynamics the corresponding fidelities decay at an effective multi-qubit $T_1$ rate and a rate that is faster than the multi-qubit $T_1$ or $T_2$ alone, respectively. In numerics, the substep decomposition of the algorithm caused these rates to partially mix, though the further effect of decomposing to native gates was smaller. In analytics, the intermediate case of diffusive but unentangled dynamics was used to clarify that the difference between the effects of dephasing and of random entanglement on fidelity decay is, in our simple model, due only to the greater influence from relaxation in the latter.}

{The gate-based Lindblad model was used with only single-qubit $T_1$ and $T_2$ noise for the single gate with the largest error, \code{CNOT}. When fit to experimental data, this model extracts far less information than full process tomography, but more than fitting a depolarizing noise model. A natural next extension of this model would be to consider multiple gates, such as by including the identity gate which can have high error due to its long duration in some circuits. The physical motivation behind the gate-based Lindblad model makes it ideal for capturing the effects of dynamics at minimal experimental cost.}

{Treating the experiment as a benchmarking exercise, first the average error per \code{CNOT} gate was estimated. Due to the fidelity dependence on dynamics, the average error varies by a factor of $1.5\times$ over the dynamics. The observed error is $3.0-4.5\times$ greater than reported by IBM-Q's method of randomized benchmarking with a depolarizing noise model. This discrepancy is largely explainable by RB being less sensitive to certain noise sources such as coherent error and low-frequency noise, and by an increase in crosstalk error when using three qubits rather than two.} However, how much the different dynamics between RB and the QSM also contribute is not yet clear.

{The noise models were also fit to experiment to extract effective $T_1$ and $T_2$ times for the \code{CNOT} gate.} All models agree well on the effective $T_2$ time, which is dominant, but diverge on the $T_1$ time, which has a weaker effect and is difficult to resolve. The effective \code{CNOT} $T_2$ time is $2.7\times$ shorter than the {$T_{2 \text{Echo}}$ of an idle qubit, suggesting \code{CNOT} error is no more than $\sim3\times$ larger than idle gate error when the noise is dominated by $T_2$.}


Retrieving useful experimental fidelities for complex Hamiltonian simulation was enabled by the highly efficient gate decomposition of the quantum sawtooth map (QSM). It combines the exactness of quantum maps, which do not require Trotterization \newtext{due to the exact decomposition of the delta-kicked potential}, with a \newtext{polynomial-length} algorithm for low-order polynomial Hamiltonians \newtext{(see Appendix \ref{appendix:gate_decomp}).} On IBM-Q the QSM required only 33 \code{CNOT} gates per evolution time step and 66 \code{CNOT} gates per {Loschmidt echo} time step. This enabled seven time steps of clearly localized dynamics and five time steps of {Loschmidt echo} fidelity decay.

A model of low-frequency parametric noise was also studied {in Appendix~\ref{appendix:parameter_noise}}  that is relevant to future experiments on more qubits. This model was previously used to show a minimum of six qubits is required to observe the Lyapunov exponent in the QSM. Here it was combined with Lindblad noise to show that the Lyapunov exponent can still be observed, but will require seven or more qubits depending on the strength of Lindblad noise. \newtext{The parametric noise model can be applied to determine the Lyapunov exponent on future fault-tolerant error-corrected quantum computers.  However, ensuring that the Lyapunov algorithm works on NISQ hardware will require the ability to control and tune the magnitude of the various types of noise processes appropriately.}

The existence of a large fidelity gap between different dynamics in digital quantum simulations also raises the question of whether arbitrary algorithms may have dynamics that affect their fidelity. In particular, the degrees of entanglement, superposition, and randomness during an algorithm could suggest particular fidelity dependence on the effective $T_1$ and $T_2$. This could be crucial for anticipating quantum advantage by better extrapolating limited device characterization to predict the fidelities of NISQ algorithms.

\section*{Acknowledgements}

The authors thank the Quantum Leap group at LLNL for stimulating discussions and ideas that improved the manuscript, including Vasily Geyko, Frank R. Graziani, Stephen B. Libby, and Yuan Shi. We also thank Jonathan L. DuBois, Kristin M. Beck, Alessandro R. Castelli, and Yaniv J. Rosen of the Quantum Coherent Device Physics group at LLNL, Kyle A. Wendt of the Nuclear Data and Theory group at LLNL, and Robert Tyler Sutherland at Oxford Ionics for their insights into quantum hardware and noise processes.

\section*{Funding}

Work for LLNL-JRNL-835805 was performed by LLNL under the auspices of the U. S. DOE under Contract DE-AC52-07NA27344 and was supported by the DOE Office of Fusion Energy Sciences “Quantum Leap for Fusion Energy Sciences” project FWP-SCW1680 and by LLNL Laboratory Directed Research and Development project 19-FS-078.

\section*{Declaration of Interests}

The authors report no conflict of interest.

\bibliographystyle{apalike}

\bibliography{qsm}

%
%

\appendix

\section{Implementation}
\label{appendix:optimize_IBM}

\subsection{IBM-Q Implementation} 

\subsubsection{Circuit optimization}

Despite the relative simplicity of the QSM algorithm, executing the circuit in Eq.~\ref{eq:QSM_decomp_gates} on IBM-Q's state-of-the-art hardware quickly accumulates error. Optimizing the circuit to reduce error can greatly improve the results.

We only attempt to reduce error by decreasing the gate count of the two-qubit \code{CNOT} gate, the largest-error gate. This is motivated by the goal of benchmarking, which would be complicated by the use of other noise mitigation techniques such as dynamical decoupling. The focus on \code{CNOT} gates is motivated by the small total contribution of single-qubit gate errors. According to IBM-Q's reported RB error estimates, each \code{CNOT} gate produces about $30\times$ more error than physical single-qubit gates, namely \code{SX} (square-root-of-not) and \code{X} gates. (\code{Z} gates are virtual with presumably even smaller error \citep{mckay2017efficient}.) This can partially be attributed to the gate duration, with a \code{CNOT} gate taking on average about $10\times$ longer to execute than \code{SX} and \code{X} gates, which have identical duration. Combining the observations that single-qubit gate error is $1/30$ of the \code{CNOT} gate error and that from table \ref{table:gate_counts} single-qubit gates appear $1/2$ as often in the circuit for \code{ibmq\_manila} $n=3$, it appears that single-qubit gates contribute no more than $2\%$ of the total error in this context. This remains true for increasing qubit count $n$ per the faster scaling of two-qubit gates in Eq.~\ref{eq:QSM_decomp_gates} and the gate decomposition in the next paragraph.

To determine the base \code{CNOT} gate count of the circuit Eq.~\ref{eq:QSM_decomp_gates} on IBM-Q, first logic gates are decomposed to native hardware gates. The simple relations for decomposing are: \code{PHASE($\phi$)} = \code{RZ($\phi$)} up to an unimportant global phase, where \code{RZ($\phi$)} is a Z-rotation by angle $\phi$; \code{H} = \code{RZ($\pi/2$) SX RZ($\pi/2$)}; $\code{SWAP}_{01}$ = $\code{CNOT}_{01} \code{CNOT}_{10} \code{CNOT}_{01}$; and lastly the \code{CPHASE} gate is given by $\code{CPHASE}_{01}(\phi) = \code{RZ}_1(\phi/2) \code{CNOT}_{01} \code{RZ}_1(2\pi-\phi/2) \code{CNOT}_{01}  \code{RZ}_0(\phi/2)$. These decompositions are performed automatically by Qiskit's \code{transpile} function.

Reducing the \code{CNOT} gate count is done with the gate count optimizer that is automatically applied by the \code{transpile} function. This does not guarantee optimal results but provides a useful improvement. For this the barriers in Fig.~\ref{fig:circuit_QSM} are removed, though barriers between map steps are retained for scalability with number of map steps. Using the above decompositions of each gate on Fig.~\ref{fig:circuit_QSM}, the base \code{CNOT} count on three fully-connected qubits is $12 \times 2 = 24$ \code{CNOT} gates. Using \code{transpile} per Appendix \ref{appendix:optimize_IBM} applies gate transformations stochastically to search for gate reductions. Out of 100 attempts the best optimization reduces gate count from $24 \rightarrow 19$ \code{CNOT}s, for a $21\%$ reduction. Smaller gate counts were found, but were for incorrect circuits due to a bug described in Appendix \ref{appendix:optimize_IBM}, or used approximations that did not generalize to other values of $\kparam$.

Retaining consistent gate count across $\kparam$ is crucial to isolating the effect of dynamics on fidelity decay. This required avoiding specific values of $\kparam$, such as $2.0$ and $4.5$, and avoiding $\kparam \ll 2\pi/\beta^2$ where some \code{CP} gates in $U_\text{pot}$ approach identity and the transpiler may eliminate them from the circuit.

Sparse device connectivity increases gate counts, which allows for further gate count optimization. For two unconnected qubits, a $\code{CNOT}_{02}$ gate is replaced with $\code{SWAP}_{01} \code{CNOT}_{12} \code{SWAP}_{01}$, transforming one gate to seven. However each $\code{CPHASE}_{02}$ gate only needs one pair of \code{SWAP}s despite its base decomposition to two \code{CNOT} gates. So Fig.~\ref{fig:circuit_QSM} before optimization transforms due to linear connectivity from $24 \rightarrow 48$ \code{CNOT} gates. Gate optimization over 100 \code{transpile} attempts reduces from $48 \rightarrow 33$ \code{CNOT}s, a $31\%$ reduction. The connectivity-caused \code{CNOT}s have partially canceled with algorithmic \code{CNOT}s, reducing the cost of sparse connectivity.

\subsubsection{Circuit scaling}
\label{appendix:circuit_scaling}

The scaling of circuit error is not very relevant to the three-qubit experiments in this paper, but it is useful for estimating the feasibility of larger simulations, e.g. that could be used to observe the Lyapunov exponent \citep{porter2022observability}. It is discussed here for completeness in understanding the QSM system as a future benchmarking tool. Gate error is assumed constant for increasing number of qubits, without accounting for error sources like crosstalk which are known to increase the average error per \code{CNOT} gate \citep{mckay2019three}. Moreover, additional scaling may depend on the device under consideration.

When scaling to more than three qubits, performing \code{CNOT} gates in parallel becomes a crucial tool in reducing total error. The degree to which the total circuit error depends on gate \textit{count} versus gate \textit{depth} of parallel gates depends on how the error of a single \code{CNOT} gate $\epsilon_\text{CNOT}$ compares to idle error of the same duration $\epsilon_\text{IDLE}$. If $\epsilon_\text{CNOT} = \epsilon_\text{IDLE}$, then performing two \code{CNOT}s in parallel has identical error as performing a single \code{CNOT} with the other qubits idle, meaning performing gates in parallel is a big gain and gate depth determines circuit error. If $\epsilon_\text{CNOT} \gg \epsilon_\text{IDLE}$, then two \code{CNOT}s in parallel have similar error as two in series, meaning parallelization does little and gate count determines circuit error. The model in Sec.~\ref{sec:lindblad_gate-based} assumed $\epsilon_\text{CNOT} \gg \epsilon_\text{IDLE}$ to determine gate-specific error to lowest order. When analyzing circuit scaling, different hypotheses about this relation can lead to a range of results. This combines with uncertainty about how parallelized a circuit is for a total effectiveness of parallelization. This effectiveness reduces error relative to a serial circuit by a factor between one and $n/2$. The relationship between $\epsilon_\text{CNOT}$ and $\epsilon_\text{IDLE}$ is experimentally estimated in Sec.~\ref{sec:experiment_fit_lindblad}.

To predict the scaling of circuit error due to \code{CNOT} gate count and gate depth to many qubits, the previous insights can be combined. Base \code{CNOT} gate count from Eq.~\ref{eq:QSM_decomp_gates} scales as $O(n^2)$ for $n$ qubits. Worse-case device connectivity (linear) increases this, as the average \code{CNOT} requires an extra $O(n)$ \code{SWAP}s, each requiring three \code{CNOT}s, to connect qubits separated by an average of $O(n)$ connections. Gate count optimization, e.g. using the \code{transpile} function or similar,  may reduce this by up to $O(n)$. Parallelization effectiveness decreases circuit error by between $O(1)$ and $O(n)$ regardless of connectivity. In summary, a circuit with all-to-all connectivity has between $O(n)$ and $O(n^2)$ circuit error scaling due to \code{CNOT} gates, and one with linear connectivity has between $O(n)$ and $O(n^3)$ scaling. These ranges depend on the effectiveness of gate count optimization and parallelization. They could be refined by using IBM's transpiler to see how well \code{CNOT} gates parallelize in the QSM across different topologies.

For the gate count optimization performed by \code{transpile}, it is unknown whether an optimal solution can be found in a scalable fashion. Finding an optimal gate reduction on all-to-all qubit connectivity is an NP-complete problem \citep{botea2018complexity}. In fact even for few qubits with linear connectivity \code{transpile} struggles to find optimal results. In table~\ref{table:gate_counts} this is demonstrated by comparing the \code{CNOT} gate count from the known algorithm Eq.~\ref{eq:QSM_decomp_gates} plus optimization in column (b) to a naive approach of compiling the raw unitary $\Uqsm$ directly in column (a). In each case about one hundred iterations of \code{transpile} is attempted. For $n = 3$ the naive approach yields over twice the gate count. Whether this suboptimal optimization can provide a scalable benefit is unclear, though focusing on optimizing few-qubit sub-circuits in large algorithms may be the most practical use case.

Since these scalings are all polynomial, they would all achieve an exponential speedup relative to the $O(n 2^n)$ operations required by a classical computer using the fast Fourier transform \citep{benenti2004quantum}. However one should keep in mind the scaling of the measurement step, since measuring the entire final state would take an exponential number of runs and destroy the speedup. The speedup is preserved only for measuring collective properties like the Lyapunov exponent \citep{porter2022observability}, localization length, or anamolous diffusion exponent \citep{benenti2004quantum}.

\subsubsection{Choosing a device}

At the time of writing IBM-Q freely provides six five-qubit devices, each containing only linear three-qubit sublattices. Among these we choose to test \code{ibmq\_manila} based on its small reported gate error (often $<6.5\mathrm{E}{-3}$ on some three-qubit sublattice) and large quantum volume (32). The now-retired five-qubit device (\code{ibmq\_5\_yorktown}) had a bow-tie shape with two triangular sublattices but much larger gate error ($>2.0\mathrm{E}{-2}$), so we performed experiments on it to explore the trade off between connectivity and gate error. In table~\ref{table:gate_counts} \code{ibmq\_5\_yorktown} and \code{ibmq\_manila} are compared, albeit 1.5 years apart, with the latter giving much better performance. Our experiments on \code{ibmq\_santiago} contemporary to the \code{ibmq\_5\_yorktown} experiments gave similar results as the more recent \code{ibmq\_manila}, indicating that the time lapse has not had a large effect. A similar result comparing \code{ibmq\_5\_yorktown} and \code{ibmq\_santiago} was also found in Ref.~\citep{pizzamiglio2021dynamical}.


\subsection{Optimizing performance on IBM-Q} 

Despite using an efficient algorithm, various difficulties can harm performance. Here are some insights for avoiding pitfalls:

\begin{itemize}
	
	\item Optimize for qubit connectivity: Circuits that are transpiled to devices with sparse connectivity will add \code{SWAP} gates to connect distant qubits. This allows for partial cancellation between the algorithm \code{CNOT} gates and the swapping \code{CNOT} gates. The automated approach to this is to use a stochastic transpiler as offered by Qiskit. To enable this level of optimization, first set the \code{transpile} function's option \code{optimization\_level} from \code{=1} (the default) to \code{=3} (the maximum). Next perform a trial-and-error optimization over the transpiler seed by iterating over the option \code{seed\_transpiler=seed} and looking for the lowest resulting gate count of \code{CNOT}s using \code{circ.count\_ops()} on the transpiled circuit. Check that the transpiled circuit produces the expected outcome in a noiseless Aer simulator, due to the bug discussed below. Consider saving the result as seed behavior may change. This procedure does not directly scale to many qubits, but on few-qubit subspaces it may be useful at NISQ scales, and may eventually be automated for real applications.
	
	\item Optimize algorithms: Look for small optimizations that can be done by hand. For example the QFT algorithm and its inverse require \code{SWAP} gates to reverse the qubit ordering. But this can be canceled with the argument \code{do\_swaps=False} and replaced with a free, manual reversal of qubit ordering for gates in between the QFT pair.

	\item Optimize coherence times: Noise properties of backend devices can drift and worsen from their reported values. IBM-Q routinely recalibrates their systems, adjusting qubit and gate pulse properties to reduce error. Each device is currently recalibrated between hourly and weekly, with time since last calibration reported on each device's status screen online. Timing experiments to occur soon after calibration can help achieve optimal performance for devices that calibrate less than hourly. Choosing the connected set of qubits with the lowest-error gates is also important and should be repeated after a calibration that re-characterizes them, though these occur less often.
	
	{
		\item Watch out for unexpected changes when concatenating circuits: A hidden pitfall (at the time of this publication) can occur when a circuit without the final measurement step is transpiled. The final qubit identities can be automatically swapped, despite that no measurement is present to absorb this change \citep{porter2021quantum}. An unsuspecting user may want to repeat an algorithm step, as done in this paper, and hope to get high gate efficiency at low cost by transpiling one step and repeating it. However, one should check whether qubit swapping has occurred and either try a new seed or undo the swaps manually to ensure there are none between steps. 
	}

\end{itemize}

\section{Gate decomposition of polynomial Hamiltonian evolution}
\label{appendix:gate_decomp}

The first step in converting the operators of Eq.~\ref{eq:QSM_def} into a series of common two-qubit gates is knowing how to implement $\hat p$ or $\hat q$ (which have identical form in their respective bases). Representing a single-body {\it state} of quantum number $\pindex=0,...,N-1$ on $n=\log(N)$ qubits can be accomplished via the binary representation \citep{georgeot2001exponential}
\begin{equation}
	\ket{p} \rightarrow \pindex=\sum_{j=0}^{n-1}\alpha_j 2^j \rightarrow \ket{\alpha_{n-1} ... \alpha_0}
\end{equation}
where $\alpha_j=0$ or $1$ for each qubit $j$ depending on the value of $p$. One can express \textit{diagonal operators} similarly. The operator $\Up=e^{-i \Tparam \hat \pindex^2/2}$ can be expressed in the $p$ basis by noting
\begin{align}
	\pindex^2 &=\sum_{j_1, j_2} \alpha_{j_1} \alpha_{j_2} 2^{j_1+j_2} \text{\quad and so} \\
	e^{-i \Tparam \pindex^2/2} &=\prod_{j_1, j_2} e^{-i \Tparam \alpha_{j_1} \alpha_{j_2} 2^{j_1+j_2-1}} \nonumber \\
	&= \prod_{j_1} e^{-i \Tparam \alpha_{j_1}^2 2^{2 j_1-1}} \prod_{j_1 < j_2} e^{-i \Tparam \alpha_{j_1} \alpha_{j_2} 2^{j_1+j_2}} \\
	e^{-i \Tparam \hat \pindex^2/2} &\rightarrow
	\left[
	\begin{pmatrix}
		1 &  \\
		&  e^{-i \Tparam 2^{2n-3}}
	\end{pmatrix}
	\otimes \dots 
	\begin{pmatrix}
		1 &  \\
		&  e^{-i \Tparam 2^{-1}}
	\end{pmatrix}
	\right] \times \nonumber \\
	&\quad\,\,\, \left[
	\bigotimes^{n-2} I \otimes
	\begin{pmatrix}
		1 &  & & \\
		&  1 & & \\
		&  & 1 & \\
		&  & & e^{-i \Tparam 2^{1}}
	\end{pmatrix}
	\right] \times \dots.
\end{align}
This describes \code{PHASE} gates on every qubit $j_1$ and \code{CPHASE} gates on every pair of qubits $j_1 < j_2$ with phase $-\Tparam 2^{j_1 + j_2 }$ (after merging $j_1>j_2$ into $j_1<j_2$). Since the desired operator $\Ukin$ is shifted by $N/2$ from $\Up$, because the correct range is $p'=-N/2, ..., (N-1)/2$, it gets modified to
\begin{flalign}
	\label{eq:Ukin}
	\Ukin \equiv \Uphase(\Tparam) &= e^{-i \Tparam (\hat \pindex-N/2)^2/2} \nonumber \\
	&\sim e^{-i \Tparam \hat \pindex^2/2} e^{i \Tparam N \hat \pindex/2}
\end{flalign}
with unobservable global phase neglected. This contributes further \code{PHASE} gates described by 
\begin{equation}
	e^{i \Tparam N \hat \pindex/2}=\prod_j e^{i \Tparam N \alpha_j 2^{j-1}}.
\end{equation}
This describes the $\Ukin$ gate, but the $\Upot$ gate is identical in its own basis with $\Tparam$ replaced by $-k \beta^2$ and $\hat \pindex$ by $\hat \qindex$ with the same $q=0,... ,N-1$:
\begin{flalign}
	\label{eq:Upot}
	\Upot \equiv \Uphase(-k \beta^2) &= e^{i k \beta^2 (\hat \qindex-N/2)^2/2} \nonumber \\
	&\sim e^{i k \beta^2 \hat \qindex^2/2} e^{-i k \beta^2 N \hat \qindex/2}.
\end{flalign}
Switching between bases as in Eq.~\ref{eq:QSM_decomp_simple} is needed to execute both $\Ukin$ and $\Upot$ with this efficient decomposition. The full algorithm is converted to gates in Eq.~\ref{eq:QSM_decomp_gates}.

\section{Fidelity decay of localized and diffusive dynamics under single-qubit Lindblad noise}
\label{appendix:Lindblad_and_dynamics}

\subsection{Fidelity of pure states}

The quantum sawooth map exhibits dynamics that range from localized to diffusive (chaotic). To study its interaction with single-qubit Lindblad noise from a theoretical standpoint, we can make some reasonable simplifying assumptions. The main assumption is that decay of an evolving density matrix is approximately the decay of a pure state with distribution and entanglement typical of the evolving state. In the localized limit this is exactly true. In the diffusive case this requires that the initial evolution to a typical diffusive state is fast relative to Lindblad decay and that after a typical diffusive state is reached the ongoing Hamiltonian evolution can be neglected. The goal is to calculate the fidelity of the decaying state relative to the initial pure state over time. The pure state assumption allows a simplified form. First the general fidelity of corrupted density matrix $\matsigma$ relative to pure state density matrix $\matrho$ is
\begin{flalign}
	f = \left( \Tr \sqrt{\sqrt{\matrho} \matsigma \sqrt{\matrho}} \right)^2.
\end{flalign}

But a pure state has the nice property that
\begin{flalign}
	\matrho^2 &= \ket{\psi} \braket{\psi} \bra{\psi} = \matrho = \sqrt{\matrho}
\end{flalign}

which simplifies the fidelity to
\begin{flalign}
	\label{eq:apx_fid_formula_as_elems}
	f &= \bra{\psi} \matsigma \ket{\psi} \nonumber \\
	&= \sum_{i,j} \sigma_{i,j} \rho_{j,i} = \sum_{i,j} \sigma_{i,j} \rho_{i,j}^* ,
\end{flalign}

the Frobenius inner product of the density matrices. It will be useful to consider the elements of the Hadamard element-wise product 
\begin{flalign}
	\label{eq:sig_dot_rho}
	(\matsigma \circ \matrho^*)_{ij} = \sigma_{i,j} \rho_{i,j}^*
\end{flalign}

which has elements that sum to the fidelity.

\subsection{Localized dynamics}

The quantum sawtooth map has a limit of localized dynamics at $\kparam \rightarrow 0$. In this limit the Hamiltonian becomes a diagonal phase matrix $\matH = \text{diag}(H_0 \, H_1)$. This only evolves the off-diagonal elements $\rho_{i,j}$ of the density matrix $\matrho$, as seen on a single-qubit example
\begin{flalign}
	[\matH,\matrho] &= 
	\begin{pmatrix}
		0 & (H_0-H_1) \rho_{01} \\
		(H_1-H_0) \rho_{10} & 0
	\end{pmatrix}.
\end{flalign}
If we assume an initial state in the computational basis $\matrho = \text{diag}(0 ... 1 ... 0)$ then the Hamiltonian evolution has no effect on $\matrho$.

\subsubsection{Single-qubit decomposition}

The effect of Lindblad noise on the localized state can be trivially decomposed into single-qubit effects, since both the Lindblad operators and density matrix decompose. 

The Lindblad equation Eq.~\ref{eq:lindblad_eq} in the main text describes the noisy time evolution of initial state $\matrho$. {Here we consider only single-qubit Lindblad collapse operators
\begin{flalign}
	\label{lindblad_L1}
	\matL_{1,j} &= I \otimes \dots
	\begin{pmatrix}
		0 & 1 \\
		0 & 0
	\end{pmatrix}_j
	\dots \otimes I \\
	\label{lindblad_L2}
	\matL_{2,j} &=  I \otimes \dots
	\begin{pmatrix}
		0 & 0 \\
		0 & 1
	\end{pmatrix}_j
	\dots \otimes I
\end{flalign}
which describe relaxation at a rate $\nu_1$ and pure dephasing \citep{krantz2019quantum} at a rate $\nu_2$, respectively, on each qubit $j$.} For a single unentangled qubit, the solution $\matsigma$ to the Lindblad equation when $H$ has no effect evolves as
\begin{flalign}
	\label{eq:apx_lindblad_evol_1Q}
	\dot \matsigma = 
	\begin{pmatrix}
		\nu_1 \sigma_{11} & -\frac{\nu_1 + \nu_2}{2} \sigma_{01} \\
		-\frac{\nu_1 + \nu_2}{2} \sigma_{10} & -\nu_1 \sigma_{11}
	\end{pmatrix}.
\end{flalign}
with decay rates that obey $\nu_1 \equiv 1/T_1$, $\nu_2 \equiv 1/T_\varphi$, and $\nu_1/2+\nu_2/2 = 1/T_2$. Solving this differential equation is simple, as three of the elements depend only on themselves, yielding exponential decays, and the $\sigma_{00}$ element is determined by trace preservation $\Tr(\matsigma) = 1$. \newtext{Trace preservation is equivalent to the conservation of probability.}

Localization simplifies both $\matsigma$ and the fidelity calculation. If the qubit starts in $\ket{\psi}=\ket{1}$, then
\begin{flalign}
	\matsigma = 
	\begin{pmatrix}
		1 - e^{-\nu_1 t} & 0 \\
		0 & e^{-\nu_1 t}
	\end{pmatrix}
\end{flalign}
and $f = \sigma_{11} = e^{-\nu_1 t}$. If it starts in $\ket{\psi}=\ket{0}$ then $f=1$. No off-diagonal terms means no $\nu_2$ dependence.

\subsubsection{Average fidelity}

It is useful to average the localized fidelity over all initial conditions for comparison to the diffusive case later. The fidelity for $n$ qubits decomposes into the product of single-qubit fidelities since both $\matrho$ and $\matL_{i,j}$ decompose. $j$ excited qubits then decay at an exponential rate $-j \nu_1$. Averaging over initial conditions includes $n$ states decaying at $-\nu_1$, ($n$ choose 2) states decaying at $-2\nu_1$, and so on. These sum to
\begin{flalign}
	\label{eq:apx_fid_loc}
	f_{ave} &= \frac{1}{2^n} \sum_{j=0}^n {n \choose j} e^{-j \nu_1 t} = \left( \frac{1 + e^{-\nu_1 t}}{2} \right)^n \nonumber \\
	&\approx 1 - \frac{n}{2} \nu_1 t + O(n^2 \nu_1^2 t^2) \approx e^{-n(\nu_1/2) t}.
\end{flalign}
The single-qubit decay rate is seen to be $-\nu_1/2$, as expected.

\subsection{Diffusive dynamics}

In the other limit of the quantum sawtooth map, large $\kparam$ leads to chaotic, diffusive dynamics. Hamiltonian evolution now has an effect, but will be simplified by two assumptions.

The first assumption is that the fast Hamiltonian evolution reaches a fully diffused state quickly enough to be considered immediate, so that the analysis begins with a randomly entangled state. To justify this, there are two relevant time scales for the initial wave function spreading: the Ehrenfest time $\tau_E$ and the Heisenberg time $\tau_H$ (see Sec.~\ref{sec:localization}). The Ehrenfest time is how long the quantum dynamics closely match the classical dynamics \citep{shepelyansky2020ehrenfest}. For a chaotic system this implies rapid spreading at the rate $e^{\lambda t}$. This must end after reaching the finite system size $N$. This implies an Ehrenfest time of $\tau_E \sim \ln(N)/\lambda$. To avoid localization $\Kparam \gtrsim N^{-2/5}$ when $L=1$, implying $\lambda \gtrsim N^{-1/5}$ and a bound on Ehrenfest time of $\tau_E \lesssim \ln(N) N^{1/5}$. For small system sizes this is clearly small, and simulations suggest $\tau_E \sim 1$.

Between the Ehrenfest time and Heisenberg time a much slower diffusion occurs. But after the Heisenberg time the effects of quantum interference are fully realized and dynamical localization occurs, halting diffusion. The Heisenberg time for localization is $\tau_H \sim \ell$ \citep{benenti2004quantum,shepelyansky2020ehrenfest}, which if extended to the diffusive regime suggests $\tau_H \gtrsim N$. Since the Ehrenfest time represents the bulk of the spreading and occurs on a short time scale, this justifies the assumption of an immediate fully diffused state.

The second assumption is that Hamiltonian evolution after full diffusion has little effect on the fidelity decay and can be neglected to simplify the Lindblad evolution. This assumption is supported by numerical simulation results in Fig.~\ref{fig:theory_vs_sim}.

A ``typical" diffused state will be considered, with random features to be averaged over. This diffused state is statistically independent of the initial state before Hamiltonian evolution, so an explicit averaging over initial conditions will not be needed. While both amplitudes and phases should be random, a further simplification is to assume that the amplitudes are nearly uniform after the diffusive process occurs, leaving only random phases. This assumption is justified by ergodicity of the dynamics. 

After these simplifications, a diffusive state is characterized by two features: statistically uniform probability distribution and random entanglement.

\subsubsection{Unentangled state}

To test uniform probability and random entanglement independently, first consider a uniform state that is not entangled. Such a state is constructed by initializing in the ground state and applying a Hadmard gate to each qubit. In this case one can again decompose to single-qubit evolution Eq.~\ref{eq:apx_lindblad_evol_1Q}, but acting on initial state
\begin{flalign}
	\ket{\psi} &= \frac{1}{\sqrt{2}} ( \ket{0} + \ket{1} ), \qquad	\matrho = \frac{1}{2}
	\begin{pmatrix}
		1 & 1 \\
		1 & 1
	\end{pmatrix}.
\end{flalign}
The Lindblad solution for initial condition $\matrho$ is
\begin{flalign}
	\label{eq:apx_sigma_single_qubit}
	\matsigma = \frac{1}{2}
	\begin{pmatrix}
		2 - e^{-\nu_1 t} & e^{-\frac{\nu_1 + \nu_2}{2} t} \\
		e^{-\frac{\nu_1 + \nu_2}{2} t} & e^{-\nu_1 t}
	\end{pmatrix}.
\end{flalign}
The fidelity is again the product of single-qubit fidelities, with each qubit contributing $f = \frac{1}{2} \sum_{i,j} \sigma_{i,j} = \frac{1}{2}(1 + e^{-\frac{\nu_1 + \nu_2}{2} t})$ where exponentials on the diagonal cancel. The $n$-qubit fidelity for any initial condition is
\begin{flalign}
	\label{eq:apx_fid_unentangled}
	f &= \left( \frac{1 + e^{-\frac{\nu_1 + \nu_2}{2} t}}{2} \right)^n \approx e^{-n(\nu_1/4 + \nu_2/4) t}
\end{flalign}
which is the same form as the averaged localized case Eq.~\ref{eq:apx_fid_loc}, but with $\nu_1 \rightarrow \frac{\nu_1 + \nu_2}{2}$.

\subsubsection{Entangled states}

In a \newtext{randomly} entangled diffusive state, the phases of each $n$-qubit basis state would be random, preventing decomposition to single qubits. This means
\begin{flalign}
	\label{eq:apx_entangled_pure_state}
	\ket{\psi} &= \frac{1}{\sqrt{2^n}} \Big( \ket{\dots 00} + e^{i \phi_1}  \ket{\dots 01} + e^{i \phi_2}  \ket{\dots 10} + \dots \Big) \\
	\label{eq:apx_entangled_pure_dm}
	\matrho &= \frac{1}{2^n}
	\begin{pmatrix}
		1 & e^{-i \phi_1} & e^{-i \phi_2} & \dots \\
		e^{i \phi_1} & 1 & e^{i (\phi_1 - \phi_2)}  \\
		e^{i \phi_2} & e^{i (\phi_2 - \phi_1)} & 1& \\
		\vdots & & & \ddots
	\end{pmatrix}
\end{flalign}
The fidelity $f = \sum_{i,j} \sigma_{i,j} \rho_{i,j}^*$ shows these phases cancel out at $t=0$ when $\matsigma(t=0)=\matrho$, giving $f=4^n/4^n=1$. But how does an $n$-qubit $\matsigma$ evolve? 


Similarly to Eq.~\ref{eq:apx_lindblad_evol_1Q}, the equation for a given $\dot \sigma_{i,j}$ will have a relaxation term $ -\nu_1 \sigma_{i,j}$ for each qubit that is 1 in both states $i$ and $j$, a gain term $ \nu_1 \sigma_{i+2^k,j+2^k}$ for each qubit that is 0 in both states $i$ and $j$ where $k$ is that qubit's index in the list of qubits, and a dephasing term $-\frac{\nu_1 + \nu_2}{2} \sigma_{i,j}$ for each qubit that is 0 in $i$ or $j$ but 1 in the other. With only the gain term coupling the matrix elements, and only along the same diagonal, it is straightforward if time-consuming to solve the set of equations on each diagonal of $\matsigma$. Then the elements of $\matsigma \circ \matrho^*$ can be summed to calculate the fidelity.

\subsubsection{Two qubits}

It is instructive to work out the fidelity for two qubits. The main diagonal terms have the simplest contribution, since $\sum \sigma_{i,i} \rho_{i,i} = \sum \sigma_{i,i}/2^n = 1/2^n$ due to trace preservation $\Tr(\matsigma)=1$. The precise dynamics on the diagonal are irrelevant for the diffusive fidelity, as was seen even in the unentangled state.

For off-diagonal terms, many high-index elements of $\matsigma$ do not have gain terms, making them simple exponential decays. For two qubits,
\begin{flalign}
	\dot \sigma_{2,3} &= -\frac{\nu_1+\nu_2}{2} \sigma_{2,3} - \nu_1 \sigma_{2,3} \nonumber \\
	\sigma_{2,3} &= \frac{1}{2^n} e^{i (\phi_2-\phi_3)} e^{-\frac{3 \nu_1 + \nu_2}{2} t}
\end{flalign}
where the indices correspond to binary states, i.e. $2=10_2, 3=11_2$. This matrix element describes qubit 0 in superposition and qubit 1 in state 1, producing the first and second terms above, respectively. In contrast, a low-index state depends on decay from high-index states:
\begin{flalign*}
	\dot \sigma_{0,1} &= -\frac{\nu_1+\nu_2}{2} \sigma_{0,1} + \nu_1 \sigma_{2,3}
\end{flalign*}
If we insert the ansatz $\sigma_{0,1} = A e^{at} + B e^{bt}$ into the equation, then we find

\begin{flalign*}
	\dot \sigma_{0,1} &= a(Ae^{at} + B e^{bt}) + B(b-a) e^{bt} \\
	a &= -\frac{\nu_1 + \nu_2}{2} \\
	b &= -\frac{3 \nu_1 + \nu_2}{2} \\
	B &= \frac{\nu_1}{2^n} e^{i (\phi_2-\phi_3)} \frac{1}{b-a}
	= -\frac{1}{2^n}  e^{i (\phi_2-\phi_3)}
\end{flalign*}
with initial condition
\begin{flalign*}
	\sigma_{0,1}(t=0) &= A +  B = \frac{1}{2^n} e^{-i \phi_1} \\
	A &= \frac{1}{2^n} \left( e^{-i \phi_1} + e^{i (\phi_2-\phi_3)} \right)
\end{flalign*}
giving the solution
\begin{flalign}
	\sigma_{0,1} = \frac{1}{2^n} \Big[ \left( e^{-i \phi_1} + e^{i (\phi_2-\phi_3)} \right) e^{-\frac{\nu_1 + \nu_2}{2} t} - e^{i (\phi_2-\phi_3)} e^{-\frac{3 \nu_1 + \nu_2}{2} t} \Big].
\end{flalign}
In the unentangled limit where all $\phi_i \rightarrow 0$, the fidelity contributions of the two elements $\sigma_{2,3}$ and $\sigma_{0,1}$ partially cancel, eliminating the faster decay. But with random phase entanglement  their combined contribution generally doesn't cancel, instead giving
\begin{flalign}
	\sigma_{0,1} \rho_{0,1}^* &+ \sigma_{2,3} \rho_{2,3}^* =
	\frac{1}{(2^n)^2} \Big[ \left( 1 + e^{i ( \phi_1 + \phi_2 - \phi_3)} \right) e^{-\frac{\nu_1 + \nu_2}{2} t} + \left( 1 - e^{i ( \phi_1 + \phi_2 - \phi_3)} \right) e^{-\frac{3 \nu_1 + \nu_2}{2} t} \Big].
\end{flalign}
In the fidelity this is further combined with its complex conjugate, resulting in cosines of the phases.

So the effect of random phases from diffusive entanglement does not change which fidelity decay rates are possible, but rather changes the relative weight of each term. This lesson will generalize to $n$ qubits. 

Since Eq.~\ref{eq:apx_fid_formula_as_elems} says fidelity is a sum over matrix elements of $\matsigma \circ \matrho^*$ (Eq.~\ref{eq:sig_dot_rho}), it is helpful to show this full matrix for two qubits:
\begin{flalign}
	\label{eq:apx_fid_matrix_two_qubits}
	\matsigma \circ \matrho^* = \frac{1}{4^2} 
	\begin{pmatrix}
		\begin{matrix}
			4 - 4 e^{- \nu_1 t} + \\ 
			e^{-2 \nu_1 t} 
		\end{matrix} &
		\begin{matrix}
			\left( 1 + e^{i  \phi_{123}} \right) e^{-\frac{\nu_1 + \nu_2}{2} t}  + \\
			\left( - e^{i \phi_{123}} \right) e^{-\frac{3 \nu_1 + \nu_2}{2} t}
		\end{matrix} &
		\begin{matrix}
			\left( 1 + e^{i \phi_{123}} \right) e^{-\frac{\nu_1 + \nu_2}{2} t}  + \\
			\left( - e^{i \phi_{123}} \right) e^{-\frac{3 \nu_1 + \nu_2}{2} t}
		\end{matrix} & 
		e^{-\frac{2 \nu_1 + 2 \nu_2}{2} t} \\
		c.c. & 2 e^{- \nu_1 t} - e^{-2 \nu_1 t} & e^{-\frac{\nu_1 + \nu_2}{2} t} & e^{-\frac{3 \nu_1 + \nu_2}{2} t} \\
		c.c. & c.c. & 2 e^{- \nu_1 t} - e^{-2 \nu_1 t} & e^{-\frac{3 \nu_1 + \nu_2}{2} t} \\
		c.c. & c.c. & c.c. & e^{-2 \nu_1 t}
	\end{pmatrix}
\end{flalign}
where $\phi_{123} \equiv ( \phi_1 + \phi_2 - \phi_3)$. The diagonal matrix elements only have terms for relaxation to lower states and gain from higher states. Off-diagonal terms also have dephasing due to superposition qubits. The number of superposition qubits for an element can be counted from its indices, i.e. element $(\matsigma \circ \matrho^*)_{2,3}$ measures the superposition between states $2 = 10_2$ and $3 = 11_2$, so only qubit 0 is in superposition between $0$ and $1$. This contributes one factor of $e^{-\frac{\nu_1 + \nu_2}{2} t}$, which combines with the decay term $e^{-\nu_1 t}$ from qubit 1. Off-diagonal gain terms are the only terms for which  the random phase factors of $\matsigma$ and $\matrho^*$ do not cancel, making them the only way phase dependence enters the fidelity.

\subsubsection{$n$ qubits}


An $n$-qubit average fidelity can be found based on this two-qubit example. First, when $\nu_1=0$ the coupling between elements disappears, eliminating the phase-dependent terms in fidelity. This is just the unentangled case with the solution Eq.~\ref{eq:apx_fid_unentangled}.

When $\nu_1 \neq 0$ the fidelity $f = \sum_{i,j} \sigma_{i,j} \rho_{i,j}^*$ generally depends on the phases. However we can calculate the \textit{average} fidelity over random phases $\phi_i$. The motivation for this phase averaging is twofold. First, the $2^n$ initial conditions of basis states $\{\ket{p}\}$ are averaged over, with each producing statistically independent phases. Second, even if the phases have some probability distribution with finite variance and zero mean, the number of independent phases that sum in the fidelity grows quickly with number of qubits. Two-qubit fidelity has only one independent phase $\phi_{123}$ per initial condition and time step, but three-qubit fidelity has four, each appearing with multiple decay rates. As the number of qubits grows, most decay rate coefficients sum over many independent phases, producing an additional effective averaging. Between these two averaging effects over the uniform distribution of phases, the variance of the mean should quickly shrink towards zero.

Some phase-dependent terms will always be controlled by just a few phases, specifically $\sim n$ phases rather than one due to $n$-fold symmetry in permuting qubit identitites. This could break the assumption of phase averaging. However, in the worst case a single such poorly-averaged decay rate coefficient will have fidelity contribution $\sim n/4^n$, so the effect of each will be small, as well as independent.

Assuming then that averaging over each phase gives an approximately correct total fidelity, one can use $\langle e^{i \phi_i} \rangle = 0$ to eliminate phase dependence in the fidelity. This does not reduce to the unentangled case. Off-diagonal elements of $\matsigma \circ \matrho^*$ lose their gain terms but not their relaxation and dephasing terms. This means probability flow from high to low states retains the loss from high but not the gain to low. Equivalently, all coupling terms in the differential equations for $\matsigma$ can be dropped, greatly simplifying each matrix element to a single term each. Diagonal elements are not changed by phase averaging since they have no phase dependence, so they retain their gain terms.

The general n-qubit average fidelity can be constructed from this understanding. The fidelity is
\begin{flalign}
	\label{eq:apx_fid_random_phase}
	f = \frac{1}{2^n} + \frac{1}{4^n} \sum_{k=1}^n e^{-k \frac{\nu_1 + \nu_2}{2} t} {n \choose k} 2^k \sum_{l=0}^{n-k} e^{-l \nu_1 t} {n-k \choose l}.
\end{flalign}
This construction goes as follows: First count the diagonal terms as the sum $\frac{1}{2^n}$ due to trace preservation. Then for each off-diagonal term $\sigma_{i,j}$, count the number of qubits $k$ in superposition between states $i$ and $j$. For $k$ qubits in superposition, the effects of dephasing combine for a decay rate contribution of $-k \frac{\nu_1 +\nu_2}{2}$. The number of qubit permutations with $k$ qubits in superposition is ($n$ choose $k$), and there are $2^k$ ways the superposition qubits can be flipped between matrix index contribution ordering $0,1$ and $1,0$, i.e. matrix element $0,3 = 000_2, 011_2$ and $2,1 = 010_2, 001_2$ have the same dephasing rate. These two factors count the number of matrix elements with that dephasing rate. The non-superposition qubits can each be in state 0 or 1, but phase averaging has eliminated gain to 0 while preserving relaxation from 1. This requires a sum over number of excited qubits. Elements with $l$ excited qubits have additional exponential fidelity decay at a rate $-l \nu_1$, and there are ($(n-k)$ choose $l$) permutations of these excited qubits. Lastly, since the diagonal elements must be excluded from these sums for not following the same rules after averaging, the index $k=0$ for zero superposition qubits must be removed. It is helpful to subtract this term after the fact.

One validation of Eq.~\ref{eq:apx_fid_random_phase} is that it can be easily modified to describe the unentangled case of Eq.~\ref{eq:apx_fid_unentangled}. When all phases $\phi_i$ are zero, instead of the gain terms averaging to zero they perfectly cancel the decay terms. To reflect this, remove the effect of decay in Eq.~\ref{eq:apx_fid_random_phase} via the replacement $e^{-l \nu_1 t} \rightarrow 1$. The sum over $\l$ then evaluates to $2^{n-k}$ and simplifying recovers  Eq.~\ref{eq:apx_fid_unentangled}.

The full Eq.~\ref{eq:apx_fid_random_phase} simplifies to
\begin{flalign}
	\label{eq:apx_fid_random_phase_simplified}
	f &= \frac{1}{4^n} \left( 1 + e^{- \nu_1 t} + 2 e^{- \frac{\nu_1 + \nu_2}{2} t} \right) ^n - \frac{1}{4^n} \left( 1 + e^{- \nu_1 t} \right) ^n + \frac{1}{2^n}  \\
	\label{eq:apx_fid_random_phase_1st_order}
	&\approx 1 - n \left( \nu_1/2 + \nu_2/4 \right) t + \frac{1}{2^n} n (\nu_1/2) t + O(n^2 (\nu_1/2 + \nu_2/4)^2 t^2) \\
	& \approx e^{- n \left( \nu_1/2 + \nu_2/4 \right) t}
\end{flalign}
\newtext{where both approximations assume early time $t \ll 1/\nu_1, 1/\nu_2$.} 
Comparing this averaged diffusive randomly entangled decay to the diffusive unentangled decay Eq.~\ref{eq:apx_fid_unentangled}, the effect of $\nu_1$ on the initial rate approximately doubled but the effect of $\nu_2$ remains the same. The extra $\nu_1$ contribution comes from relaxation terms that no longer cancel with gain terms.

The diffusive fidelity Eq.~\ref{eq:apx_fid_random_phase_simplified} can be understood more intuitively by noting the phase averaging decouples all matrix elements except the $n$-qubit diagonal. A single decoupled, phase-averaged qubit within the $n$-qubit system, ignoring the trace preservation from the $n$-qubit diagonal, can be considered as having a fidelity matrix $\matsigma \circ \matrho^*$ of
\begin{flalign}
	\label{eq:apx_fid_matrix_single_phase_averaged_qubit}
	 \frac{1}{4}
	\begin{pmatrix}
		1 & e^{-\frac{\nu_1 + \nu_2}{2} t} \\
		e^{-\frac{\nu_1 + \nu_2}{2} t} & e^{-\nu_1 t}
	\end{pmatrix}
\end{flalign}
where the gain term has been removed relative to Eq.~\ref{eq:apx_sigma_single_qubit}. The fidelity $f = \sum_{i,j} \sigma_{i,j} \rho_{i,j}^*$ of this to the $n$th power for $n$ qubits forms the first term in Eq.~\ref{eq:apx_fid_random_phase_simplified}. The latter two terms are corrections for the $n$-qubit diagonal, since the diagonal has no phases to average as is clear from initial state Eq.~\ref{eq:apx_entangled_pure_dm}. The second term removes the erroneous diagonal term and the third replaces it with the appropriate $1/2^n$ from trace preservation.

\section{Parametric noise}
\label{appendix:parameter_noise}

\subsection{Effects of parametric noise}

{It is worth comparing Lindblad noise models to a parametric noise model that is more common in the study of dynamics in quantum maps. This section demonstrates that the forms of the fidelity decay are very different, making this parametric noise model a poor fit to the experimental results of Sec.~\ref{sec:main_results}. However the simulations of this and the following section may be useful for explaining future experiments where low-frequency or coherent noise may dominate.}

Hamiltonian noise models have been studied widely in the quantum maps literature \citep{jacquod2001golden, benenti2001efficient, benenti2002quantum,benenti2002eigenstates,benenti2003dynamical,benenti2004quantum,frahm2004quantum,wang2004stability,gorin2006dynamics,jacquod2009decoherence}, with noise in a parameter being a common variant. Hamiltonian noise models use unitary Markovian errors representing low-frequency or coherent noise. In this section, we study the model of Ref.~\citep{benenti2002quantum,benenti2003dynamical,benenti2004quantum} in which Hamiltonian evolution is perturbed stochastically at each discrete map step by $\kparam \rightarrow \kparam + \Delta \kparam$. The noise $\Delta \kparam$ is drawn from a normal distribution with standard deviation $\sigma$, where the PDF is given by
\begin{flalign}
	p(\Delta \kparam) = \frac{1}{\sqrt{2\pi} \sigma} e^{-\frac{\Delta \kparam^2}{2 \sigma^2}}.
	\label{eq:k-noise}
\end{flalign}
This model is discussed more thoroughly in Ref.\citep{porter2022observability}. This produces a similar effect as the fixed Hamiltonian perturbation discussed in Ref.~\citep{jacquod2009decoherence}, which itself resembles static coherent noise. Coherent noise can be divided into a component that is consistent over many gates, which can often be fixed through calibration, and a component that varies, which is more difficult to address. The latter resembles parametric noise.

When this parametric noise model is used, the initial conditions $\ket{\pindex}=\ket{0},\ket{-N/2}$ are excluded from simulations because their symmetry with respect to the map causes their noise-averaged fidelity to asymptotically approach $2/N$ rather than $1/N$. This effect can be misleading on small systems so is avoided here for clarity.

In the presence of random parametric noise, the fidelity of a chaotic map initially decays as $e^{-A n_g t}$ with Fermi golden rule $A \sim \sigma^2$ for noise magnitude $\sigma$ and $n_g$ gates \citep{frahm2004quantum}, until the golden rule breaks down for large $\sigma$ \citep{porter2022observability}.  In the presence of static unitary noise the fidelity initially decays tangent to $e^{-A n_q n_g^2 t}$ with $A \sim \sigma^2$ for $n_q$ qubits, but at some fraction of the Heisenberg time this transitions to a faster Gaussian decay of $\ln(f) \sim -t^2$ \citep{frahm2004quantum}. Quantum dynamical effects, such as Lyapunov and algebraic decay described below, do not immediately set in, so they do not affect the above initial decay rates.

It was previously shown that in the diffusive regime a minimum of six qubits is needed to observe the fidelity decaying at the Lyapunov exponent rate \citep{porter2022observability}. However in the localized regime an algebraic decay rate is predicted instead \citep{jacquod2009decoherence, cucchietti2004loschmidt, porter2022observability}. This effect is shown in the QSM in Fig.~\ref{fig:sim_param_noise}(b), close to the predicted $f \propto 1/t$ at intermediate times for dimension $d=1$ and no time auto-correlation of the noise. Since algebraic decay has less stringent constraints than Lyapunov rate decay, it might be observable on fewer than six qubits, though no fewer than four as discussed below.

\changedtext{For parametric noise, $f(t=2 t_\text{fb})$ still indicates the fidelity during forward-and-back noise for $t_\text{fb}$ map step pairs, and its Fermi golden rule decay has double the decay rate relative to forward-only noise decay, that is $f_\text{FGR,fb}(t) \approx f_\text{FGR,f}(2t)$ where $t$ is the total forward-and-back time regardless of noise. But its Lyapunov rate decay depends only on the number of forward steps, so that $f_\text{Lyap,fb}(t) \approx f_\text{Lyap,f}(t) \approx e^{- \lambda t/2}$ for semiclassical Lyapunov exponent $\lambda$, showing no increased rate \citep{porter2022observability}.} A similar effect cannot be tested for in algebraic decay, since after fitting the initial fidelities $f_\text{alg}(t \gg 1) = f_0 t_0 / t$ and $f_\text{alg}(2 t \gg 1) = f_0\, 2 t_0 / 2 t$ are equivalent.

\begin{figure}
	\centering
	\includegraphics[width=60mm]{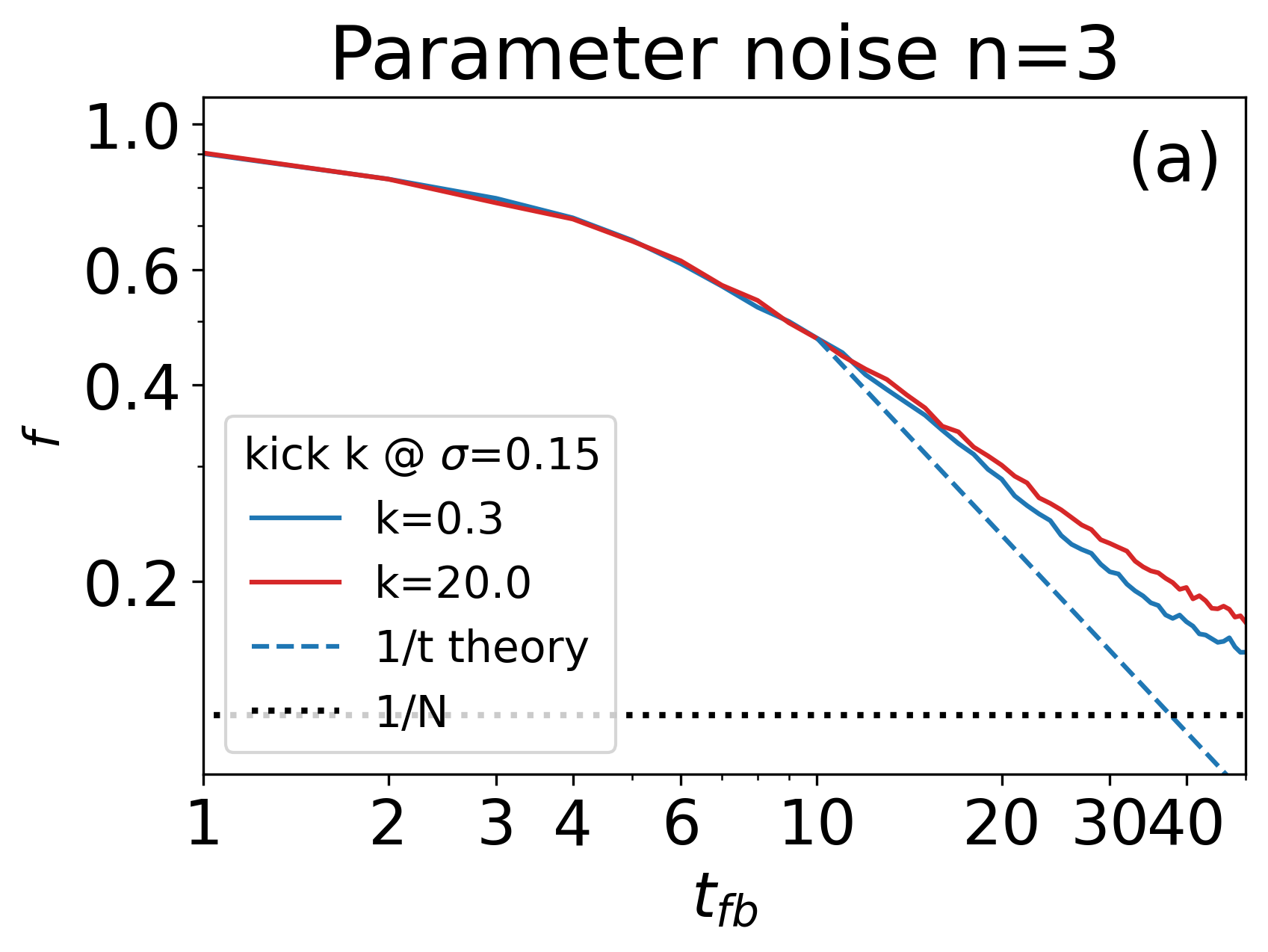}
	\includegraphics[width=63mm]{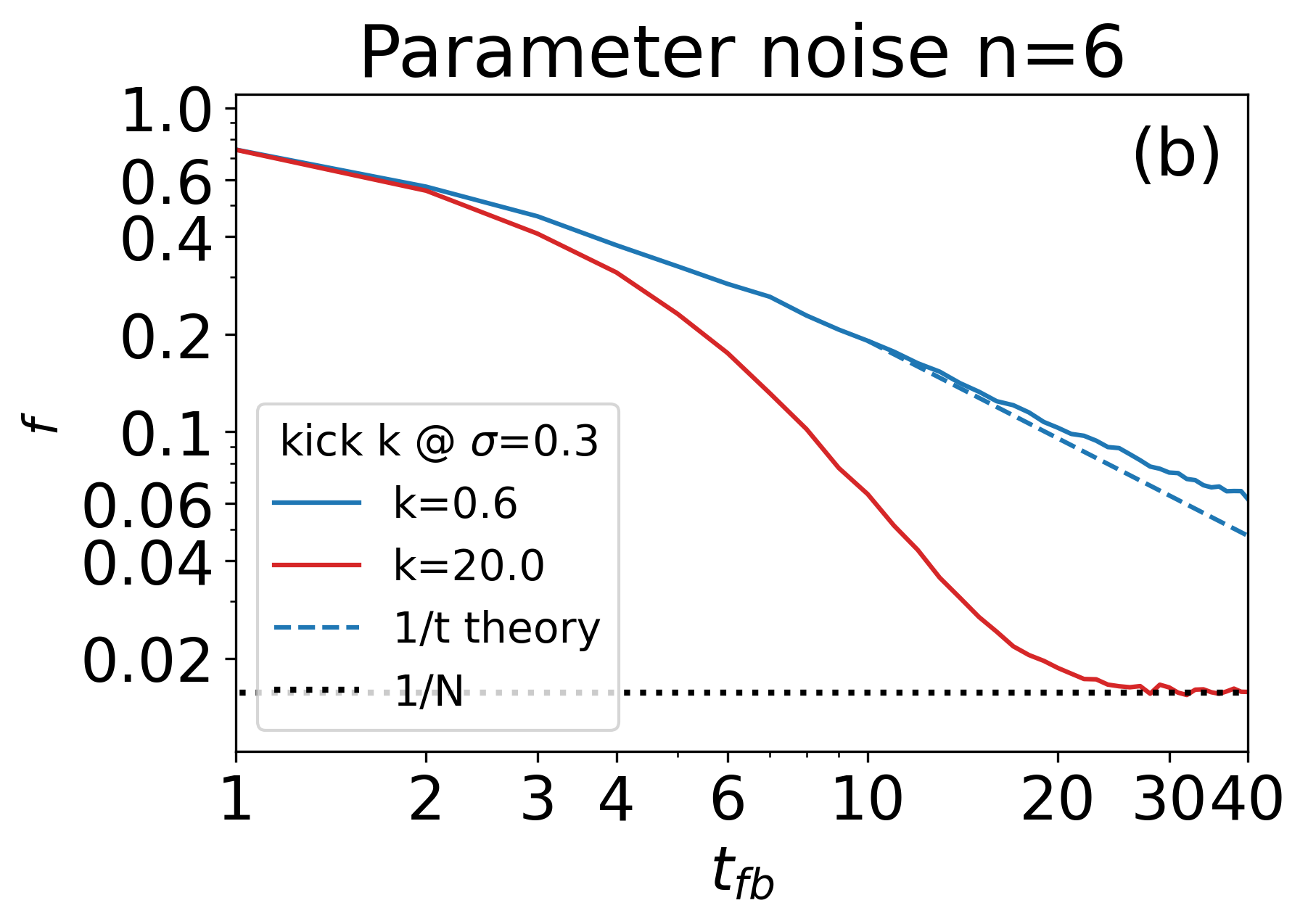}
	\caption{Parametric noise simulations for (a) n=3 and (b) n=6 on log-log plots to show the algebraic decay rate for localized dynamics. Values of the kick $\kparam$ and Gaussian noise $\sigma$ are chosen to avoid $\kparam_\text{eff}=\kparam+\Delta \kparam<0$ while keeping the number of map steps small. The resulting localized values of $\kparam$ are further from zero due to this. Using 1000 and 100 realizations of the stochastic noise plus 6 and 62 initial conditions for (a) and (b) respectively.}
	\label{fig:sim_param_noise}
\end{figure}

Since the diffusive and localized cases have different qualitative behavior under parametric noise, a fidelity gap between the two should appear. This might be confused with the effect of Lindblad noise in a sufficiently noisy experiment, so it is important to check when algebraic decay might occur. Simulations with only parametric noise show a gap is only observed on four or more qubits. On three qubits the gap is in fact reversed, with diffusive fidelity decaying slightly more slowly than localized fidelity. This is demonstrated in Fig.~\ref{fig:sim_param_noise}(a) where a reversed gap of about $3\%$ (absolute) appears. Interpreting this is complicated by the use of a Guassian parametric noise distribution which in the case shown causes $2\%$ of map steps to have $\kparam_\text{eff}=\kparam+\Delta \kparam<0$, where anomalous diffusion occurs. However the reversed gap persists when anomalous diffusion is avoided with small noise $\sigma \ll \kparam$, so that is not the main cause. A reversed gap or no gap is observed consistently when comparing $0 \leq \kparam \leq 1.9 \approx \kparam_\text{loc}$ to $\kparam=20$ at noise values $0 \leq \sigma \leq 1.0$. Comparing the decay rates to $1/t$ suggests that the localized case is still close to algebraic decay, but the diffusive case is limited by the system size to an even slower decay. This result for three qubits suggests the predictions of Lindblad noise are more relevant than parametric noise in present day experiments.

Dynamical localization displays Poisson-like energy spacing statistics, and so indicates locally integrable quantum dynamics, which is the source of the algebraic fidelity decay \citep{jacquod2009decoherence}. But a dynamical fidelity gap can also occur in locally integrable quantum systems such as Ref.~\citep{lysne2020small} (see their Fig.~3).

\begin{figure}
	\centering
	\includegraphics[width=80mm]{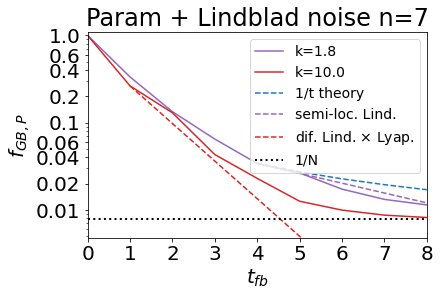}
	\caption{Comparing combined Lindblad and parametric noise simulations (solid) to theory (dashed) for the semi-localized (purple, $\kparam=1.8$) and diffusive (red, $\kparam=10.0$) cases. Theory (dashed) includes $1/t$ (blue), Lindblad semi-localized (purple) and the product of Lindblad diffusive and Lyapunov rate decays (red). {Simulations average over all 128 initial conditions at one random realization each.} All results use $n=7, \nu_1=0.025, \nu_2=0.05, \text{ and } \sigma=0.9$. For Lyapunov rate decay, theory assumes $f_\text{fb,Lyap}(t_\text{fb}) = e^{- \lambda t_\text{fb}}$ as observed.}
	\label{fig:combined_param_lindblad_noise}
\end{figure}

\subsection{Combining Lindblad and parametric noise}

The theory of Appendix \ref{appendix:Lindblad_and_dynamics} demonstrates that Lindblad noise by itself does not cause the same qualitative behavior as the parametric noise model; e.g. algebraic decay or exponential decay at the Lyapunov rate. Combining Lindblad and parametric noise can generate these effects, but if the parametric noise is small relative to the Lindblad noise the effects may be difficult to observe.

If the noise processes are statistically independent, then Lindblad and parametric noise fidelities should multiply together. The simulations of the combined noise model in Fig.~\ref{fig:combined_param_lindblad_noise} support this conclusion for both localized and chaotic dynamics. In the chaotic case the rate slows for $1 \leq t_\text{fb} \leq 4$ to the product of the Lindblad diffusive rate and the Lyapunov rate, matching pure parametric noise simulations. If the two fidelities instead summed then the slowest exponential rate, namely the Lindblad rate, would dominate by outlasting the others, but that would predict a $3\times$ slower decay than observed. The localized case also supports multiplying fidelities, as during $4 \leq t_\text{fb} \leq 7$, near the start of $1/t$ decay in Fig.~\ref{fig:sim_param_noise}, the decay remains faster than either the $1/t$ or semi-localized (Eqs.~\ref{eq:fid_gb_p_approx} and \ref{eq:rate_semi_localized}) analytic decays, rather than one rate dominating. By $t_\text{fb}=8$ the $1/N$ term begins to dominate and slow the decay.

In Ref.~\citep{porter2022observability} three bounds were derived on the observability of Lyapunov decay under parametric noise. However if dynamical and Lindblad noise do have multiplicative effects on the fidelity decay, then one can predict how this noise combination would cause these bounds to change. Bound (1) should remain unchanged, as Lindblad noise likely affects both Fermi golden rule decay and Lyapunov decay equally, causing no change in their competition. Bound (2) should also remain unchanged, as the transition to localization likely occurs at the same parameter value. But bound (3) would tighten, since it describes the limited time steps available to observe Lyapunov decay, which the addition of Lindblad noise would certainly reduce.

\section{Rationale for gate-based Lindblad model}
\label{appendix:rationale}

Working with a complete decoherence model of quantum gates is quite challenging because there are many terms that are potentially active \citep{breuer2002theory}.  For a linear trace-preserving quantum process model, the number of terms is $N^4-N^2$, where $N$ is the number of states.  Clearly, for all-to-all connectivity, a complete characterization is not scalable to many qubits.

Performing quantum process tomography \citep{chuang1997prescription} for a single entangling gate already requires a relatively large number of measurements. While there are software packages such as pyGSTi \citep{blume2013robust,nielsen2020probing} that allow one to automatically characterize a complete set of two-qubit gates, many of the commercial quantum hardware platforms available today do not allow enough run time to perform a complete characterization of even a single entangling gate. Moreover, because the complete protocol takes so long to run, the hardware characteristics may drift significantly over the duration of the experiments.

To make headway under such circumstances, one must attempt to utilize plausible simplifications of the number and types of decoherence processes under consideration. Assume that for each gate, one can identify a simplified set of decoherence processes that are dominant. One could then perform a restricted form of process tomography to characterize each gate. One could even characterize an entire gate set -- assuming that the number of independent decoherence processes is small enough that this procedure is scalable. 

Presumably, a detailed understanding of the physical processes that mediate decoherence mechanisms would yield a relatively concise mathematical description of the dominant decoherence processes. However, this could still potentially be quite complicated, for instance including non-Markovian processes, and would require custom experiments to diagnose and verify. Moreover, this could potentially require a physical model for the environment as well as for the quantum computing hardware, making the model non-Markovian and more complex than the process tomography paradigm typically admits.

Yet simply understanding the way that the physical system imposes a mathematical structure on the decoherence processes is potentially quite valuable. For example, theoretical studies have proven that the Lindblad master equation \citep{breuer2002theory} is the appropriate infinitesimal generator of the temporal evolution of a quantum system coupled to an incoherent process that is both trace-preserving and completely positive. Hence, this implies that the Lindbladian form should be considered primary and that integrating the master equation over a finite time interval should be used to determine the associated process matrix.  Clearly, a sparse representation in terms of Lindblad operators will typically yield a rather dense process matrix.  Thus, from the physical perspective, it is much more likely for the Lindblad representation to be sparse than it is for the associated process matrix.

Another plausible physical assumption is that the dominant noise processes are from one- and two-qubit interactions. This reduces the number of processes exponentially, from $O(N^4)$ (the elements of the infinitesimal process matrix) when including up to all-qubit processes to just $O(\log^2(N))$ (the number of qubit pairs) for up to two-qubit processes. In the infinitesimal process matrix this corresponds to an exponential reduction in the number of elements. This can be seen by considering the relationship between the Lindblad equation, which describes the time evolution of the density matrix, and the infinitesimal process matrix, which describes the coupling of density matrix elements. Most process matrix elements of a large system couple density matrix elements between which more than two qubits have differing states. So limiting to one- and two-qubit processes treats most process matrix elements as zero, leaving an exponentially small number of elements active.

Two-qubit processes could be further limited to fewer physically plausible channels. For instance photon exchange between qubits mediated by the environment could be an important channel, since it requires only one photon and is an extension of single-qubit relaxation and excitation processes \citep{wendtprivate}. A stronger restriction is limiting interactions to nearest neighbors in a sparsely connected device, reducing the number of processes to $O(\log(N))$. This also strengthens the assumption that no more than two-qubit processes need to be considered.

Two-qubit dephasing is a physically plausible channel that we found in our numerical studies to have very similar qualitative effects as the standard single-qubit dephasing. Presumably, the reason is that each multi-qubit dephasing process can be understood as single-qubit dephasing in an alternate basis. Understanding whether or not such non-standard dephasing terms are important for describing actual quantum hardware performance could be an interesting subject for future work.

The greatest physical simplification that can plausibly be applied to a Lindblad model is that only the single-qubit relaxation and dephasing processes are important, averaged over all qubits for simplicity, but that they are enhanced over their measured values during gate operation. (For typical quantum hardware platforms, the temperature is so far below the excitation energy that excitation processes can be considered subdominant.)  While this may be a vast oversimplification, these processes are clearly important because they are universally required to model experimental data in practice. The physical meaning of this model is simple: the physical processes that actuate the gate renormalize the interaction between the qubit and the environment, and, hence, renormalize the characteristic single qubit relaxation and dephasing rates.  Note, however, that the predictions are nontrivial because the associated time-integrated process matrix has a relatively dense form that depends on the products of the rates with the overall gate time.

While many other decoherence processes are likely needed for a truly accurate description of the quantum hardware evolution, we have found that this simplest Lindblad model is sufficient for modeling the experimental data described in this work. Our work here has the restricted goal of attempting to determine the single-qubit decoherence enhancement for the \code{CNOT} gates that clearly dominate the IBM-Q error budget. Moreover, given the limited experimental data that one is able to obtain using today’s platforms, the data may not contain enough information to robustly determine additional free parameters over and above those contained in this simple model.  In the future, it could be interesting to explore a more comprehensive approach to the determination of the relevant decoherence processes. Perhaps one could even attempt a version of gate set tomography \citep{blume2013robust} that deduces the relevant relaxation and dephasing enhancements for every qubit and every gate in the set.

\end{document}